\documentclass[%
superscriptaddress,
frontmatterverbose, 
 amsmath,amssymb,
 aps,
pra,
]{revtex4-2}
\usepackage{graphicx}
\usepackage{dcolumn}
\usepackage{bm}
\usepackage{hyperref}
\usepackage{xcolor}
\relax 
\usepackage{color}
\newcommand{\edit}[1]{{\color{black}#1}}

\begin{document}

\preprint{Propulsion and interaction of wave-propelled interfacial particles}

\title{Propulsion and interaction of wave-propelled interfacial particles}

\author{Daniel M. Harris}
 \email{daniel\_harris3@brown.edu}
\affiliation{
School of Engineering, Center for Fluid Mechanics,
Brown University, 184 Hope Street, Providence, RI 02912}

\author{Jack-William Barotta}
\affiliation{
School of Engineering, Center for Fluid Mechanics,
Brown University, 184 Hope Street, Providence, RI 02912}

\date{\today}
\maketitle

\section*{Abstract}
\textbf{When a floating body is internally or externally vibrated, its self-generated wavefield can lead to steady propulsion along the interface. In this article, we review several related and recently discovered systems that leverage this propulsion mechanism and interact hydrodynamically with one another via these surface waves.  Particles with an onboard oscillatory driver may self-propel by virtue of a fore-aft asymmetric wavefield, a phenomenon with demonstrated relevance to biological and artificial systems across scales.  Freely floating particles on a vibrated fluid bath can also self-propel along straight paths, but may also rotate in place or move along curved arcs, depending sensitively on the particle asymmetries and driving parameters.  Such surfing particles interact at a distance through their mutual capillary wavefield and exhibit a rich array of collective dynamics.  Overall, these accessible, tunable, and visually appealing systems motivate future investigations into a number of outstanding questions in fundamental fluid mechanics, while potentially also informing advances in the fields of active matter, hydrodynamic quantum analogs, and robotics.}

\section{Introduction}

While the concept of wave-driven propulsion was first proposed and realized by Longuet-Higgins in 1977 \cite{longuet1977mean}, the subject has seen increased interest over the past several years due to its relevance to a variety of natural and artificial systems.  The core physical idea underlying this propulsion mechanism is that surface waves carry net momentum, and thus floating structures experience net hydrodynamic forces whenever mediating such waves \cite{longuet1964radiation}.  In particular, a floating body capable of generating its own waves through oscillatory motion can experience a net thrust, provided the wavefield (and thus the accompanying momentum flux) is asymmetric.  This propulsion mechanism was demonstrated via self-propulsion of a mobile wavemaker by Longuet-Higgins \cite{longuet1977mean}, but more recently documented for the case of a honeybee trapped at an interface \cite{roh2019honeybees}, for human-driven oscillation of a watercraft in the technique of ``gunwale bobbing'' \cite{benham2022gunwale}, and for a number of small-scale artificial systems driven either by an onboard driver \cite{rhee2022surferbot,benham2024wave} or global oscillatory forcing \cite{ho2023capillary,barotta2023bidirectional}.  In all cases, an asymmetric kinematic motion of the floating body (or its appendages) leads to self-propulsion via self-generated surface waves.

Such wave propulsors are naturally surrounded by a periodic interfacial ``wake'' in the form of radiated capillary-gravity waves, capable of communicating dynamic stresses at a distance.  Nearby structures or other surface dwellers respond to these incident waves, facilitating rich inter-particle interactions  \cite{de2018capillary, ho2023capillary,oza2023theoretical,barotta2025synchronization,sungar2025synchronization} as well as wave-mediated self-interaction via reflected waves \cite{tarr2024probing}.  Collections of wave-driven particles exhibit novel collective behaviors and serve as a new tabletop platform for explorations of inertial active matter in fluid environments \cite{marchetti2013hydrodynamics,bechinger2016active}.  Additionally, they represent a macroscopic wave-particle duality with the potential for analogy to quantum-like behaviors \cite{bush2024perspectives} and connections to other wave-particle systems such as acoustically levitated particles \cite{lim2024acoustic}.

Our focus in the present perspective article is primarily on systems at the so-called ``capillary scale,'' an intermediate regime where surface tension, gravity, and inertia are all generally relevant.  For liquid-gas interfaces, such dynamic partnerships most commonly arise at the millimeter to centimeter scales, with motion at speeds on the order of centimeters per second.  Prior work at these scales has largely been motivated by water-walking insects \cite{bush2006walking}, who also use surface waves for propulsion in some cases, but to a lesser extent than the systems of focus here.  These remarkable insects have since inspired a vast number of bioinspired robotic counterparts for applications such as environmental monitoring \cite{yuan2012bio}.  Capillary-scale systems are generally situated in the intermediate Reynolds number regime ($Re\sim O(1-100)$) where fluid inertia is fundamental, but viscosity often plays a non-negligible secondary role.  Significantly less is known about hydrodynamic propulsion and interactions in this mesoscale regime as compared to the more well-studied low Reynolds (e.g. bacteria) and high Reynolds (e.g.  birds, fish) regimes \cite{klotsa2019above}.  \edit{Fundamental explorations of reduced models of hydrodynamic propulsion at intermediate Reynolds numbers is a topic of growing interest \cite{dombrowski2020kinematics, derr2022reciprocal,herrera2024propulsive,cobos2024spontaneous,shen2025inertia}.}

We now outline the structure of the paper. In section \ref{sec:fundamentals}, we  review the relevant fundamentals associated with interfacial flotation, translation, and capillary interactions, as well as essentials regarding capillary-gravity waves and wave radiation stress.  We then move to review the recent experimental and modeling work on wave propulsion in section \ref{sec:propulsion} and wave interaction in section \ref{sec:interaction}.  We follow this with an extensive yet non-exhaustive discussion of open questions and future directions throughout section \ref{sec:outlook}, and ultimately conclude in section \ref{sec:conclusion}.

\section{Fundamentals}\label{sec:fundamentals}
\begin{figure}
    \centering
    \includegraphics{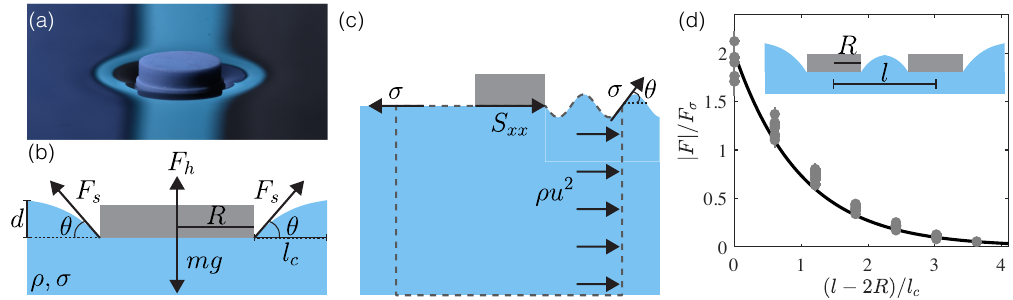}
    \caption{(a) Photo of a circular capillary disk of radius $R=0.35$ cm resting on an air-water interface. (b) Schematic of a circular capillary disk in static equilibrium. (c) Net radiation stress due to presence of waves propagating away from the right side of the body. (d) Capillary attraction for centimeter-scale capillary disks. Points are experimental measurements for disks of radius $R=0.8-1.6$ cm and mass $m=0.53-2.74$ g \cite{ho2019direct}, and the line is equation \ref{eqn:cheeriosdisk}. Inset: Schematic of the corresponding experiment where $l$ is the center-to-center distance and $l-2R$ is the edge-to-edge distance.}
    \label{fig:Fundamentals}
\end{figure}

\subsection{Static Equilibrium}\label{sec:equilibrium}
When a solid particle is gently deposited at an interface, it can remain trapped there provided hydrostatic and capillary effects can balance gravity \cite{vella2015floating}.  Much of our relevant work thus far has focused on what we refer to as ``capillary disks,'' where the fluid is in contact with only the flat base of a floating disk and the contact line remains pinned to a sharp corner along its perimeter (Figure \ref{fig:Fundamentals}(a,b)). 
This geometry avoids complications associated with contact line motion, such as contact line hysteresis and dynamic contact angles, which can be difficult to characterize and model.  

For the case of a circular capillary disk of radius $R$, the force balance in equilibrium can be expressed as 
\begin{equation}
    mg=\rho g \pi R^2 d + 2\pi R\sigma \sin\theta\label{eqn:equilibrium}
\end{equation}
where $m$ is the mass of the disk, $d$ is the displacement of the base below the equilibrium bath height, and $\theta$ is the contact angle of the fluid with the disk (Figure \ref{fig:Fundamentals}(b)) \cite{oza2023theoretical}. For small deformations, the profile of the free surface surrounding the disk ($r\geq R$) is given by the solution to the linearized Young-Laplace equation
\begin{equation}
    h(r)=-d \frac{\mathrm{K}_0(r/l_c)}{\mathrm{K}_0(R/l_c)}\label{eqn:hr}
\end{equation}
where $\mathrm{K}_0$ is the modified Bessel function of the second kind of order zero, and is analogous to a decaying exponential in a cylindrically symmetric system.  The parameter $l_c=\sqrt{\frac{\sigma}{\rho g}}$ is the capillary length, and is the characteristic length scale over which the interfacial disturbance created by the disk decays.  For a clean air-water interface the value is $l_c=0.27$ cm.  Using this result, and continuing to assume small deformations such that $\sin\theta\approx h'(R)$, 
equation \ref{eqn:equilibrium} can be rewritten as 
\begin{equation}
    mg=\pi \sigma \left(Bo + 2 \sqrt{Bo}\frac{\mathrm{K}_1(\sqrt{Bo})}{\mathrm{K}_0(\sqrt{Bo})}\right)d=k_e d\label{eqn:equilibrium2}
\end{equation}
where $Bo=\frac{\rho g R^2}{\sigma}=\frac{R^2}{l_c^2}$ is the Bond number.  Under these assumptions, the surface responds in static equilibrium as a linear spring with an effective spring constant $k_e$.  The geometric factor ${\mathrm{K}_1(\sqrt{Bo})}/{\mathrm{K}_0(\sqrt{Bo})}$ in the capillary term is always greater than 1, and approaches 1 for $Bo \gg 1$ ($R \gg l_c$) \cite{ho2019direct}.  Should the requisite contact angle $\theta$ exceed $\theta_a-\pi/2$, where $\theta_a$ is the advancing contact angle of the three-phase contact line, the capillary disk configuration is not possible.  The use of hydrophobic or super-hydrophobic materials is thus necessary. 

\subsection{Capillary-Gravity Waves}\label{sec:wavegravcap}
In order to review the core fundamentals behind the various wave phenomena discussed throughout the present article, let us begin by considering a monochromatic plane wave at an air-water interface that propagates in the positive $x$ direction of form $\eta(x,t)=A\cos\left(kx-\omega t\right)$.  Here $A$ is the wave amplitude, $k=\frac{2\pi}{\lambda}$ is the wavenumber (with $\lambda$ representing the wavelength), and $\omega$ is the wave frequency.  Assuming potential flow and small deformations, and by applying the appropriate boundary conditions, it can be shown that $\omega$ and $k$ are related to one another and the physical parameters via the dispersion relation
\begin{equation}
\omega^2=\left(gk+\frac{\sigma k^3}{\rho}\right)\tanh{kH}\label{eqn:dispersion}
\end{equation}
where $g$ is the acceleration due to gravity, $\sigma$ is the surface tension, $\rho$ is the fluid density, and $H$ is the fluid depth \cite{kundu2024fluid}.  In the case of ``deep'' water, i.e. $kH\gg1$, the expression becomes independent of depth as $\tanh{kH}\approx 1$.  In practice this criterion is not particularly stringent, as for $\tanh{kH} = 0.95$, the depth only needs to be approximately $0.3\lambda$.  We thus appeal to this limit henceforth. 

The first term in the parenthesis of equation \ref{eqn:dispersion} corresponds to the influence of gravity whereas the second term is that due to capillarity.  Their influences become of equal importance when $\rho g=\sigma k_m^2$, which for a clean air-water interface corresponds to a wavelength of $\lambda_m=2\pi l_c=1.7$ cm (and corresponding frequency of $f=\frac{\omega}{2\pi}= 13.5$ Hz).  Shorter waves (higher frequency) are more dominated by surface tension whereas longer waves (lower frequency) are more dominated by gravity.  The peaks of the propagating waveform move with the phase velocity $c(k)=\omega/k$ whose magnitude depends on the wavenumber, and has a local minimum at the wavenumber $k_m$ defined earlier.  For a clean air-water interface, this minimum phase speed is $c_m=23$ cm/s. 

These results are derived from an underlying potential flow analysis, but viscosity will inevitably be present in practice \cite{armaroli2018viscous}.  However, when viscous effects are relatively weak, the key results described thus far are essentially unchanged.  One practical consequence, however, is that the wave amplitude will decay approximately according to 
\begin{equation}
    \frac{dA}{dt}=-2\nu k^2A
\end{equation}
where $\nu$ is the kinematic viscosity.  This result suggests a characteristic viscous decay time $t_\nu=(2\nu k^2)^{-1}$.  When compared to the oscillation timescale of the waves, a dimensionless viscosity can be formed as $\epsilon=\frac{2\nu k^2}{\omega}$, which can also be interpreted as a reciprocal wave Reynolds number and is independent of the wave amplitude $A$.  For air-water waves at the minimum phase speed point, this value is $\epsilon=0.003$.  \edit{In steady state, the viscous attenuation results in an exponential decay of the wave amplitude away from a planar wave source} \cite{behroozi2001direct}.  In the weakly viscous limit ($\epsilon\ll 1$), it can be shown that the full governing equations reduce to the potential flow equations with the leading order effects of viscosity entering as modifications to the free-surface kinematic and dynamic boundary conditions \cite{dias2008theory, Milewski2015}.  This so-called ``quasi-potential'' framework has been fruitfully applied to a variety of capillary-scale free-surface problems over the past decade \cite{Milewski2015,galeano2017non,galeano2021capillary,alventosa2023inertio} including to wave propulsion and interaction \cite{oza2023theoretical,benham2024wave,barotta2025synchronization}, discussed in more depth later.

\subsection{Radiation Stress}\label{sec:radiation}

Waves, such as electromagnetic or acoustic, impart a force when impinging on a surface due to a transfer of momentum.  This phenomenon is often referred to as radiation pressure.  Capillary-gravity surface waves have an analogous effect when interacting with a floating or submerged structure, which is often referred to as radiation stress (to avoid confusion with the usual isotropic fluid pressure).  To this end, Longuet-Higgins and Stewart define radiation stress as ``the excess flow of momentum due to the presence of the waves'' \cite{longuet1964radiation}.  Longuet-Higgins and Stewart neatly outline the derivation for the radiation stress associated with a plane wave, which we briefly review here for deep water capillary-gravity waves.  

Imagine a floating raft at the center of a fluid domain in 2D that emits a propagating wave on its right side in the positive $x$ direction, described once again by $\eta(x,t)=A\cos\left(kx-\omega t\right)$, while the interface remains flat on its left side (Figure \ref{fig:Fundamentals}(c)).  \edit{According to potential flow theory, the horizontal and vertical velocities of the fluid in the wavy region are $u(x,z,t)=A\omega e^{kz}\cos\left(kx-\omega t\right)$ and $w(x,z,t)=A\omega e^{kz}\sin\left(kx-\omega t\right)$, respectively \cite{kundu2024fluid}.}  By considering horizontal momentum conservation in a control volume spanning the raft as depicted in the figure, and averaging over a cycle (denoted by the brackets $\langle\cdot\rangle=\edit{\frac{\omega}{2\pi}\int_0^{2\pi/\omega}\cdot \ dt}$), one can arrive at the following expression for the radiation stress 
\begin{equation}
    S_{x x}=\left\langle\int_{-\infty}^\eta\left(p+\rho u^2\right) d z\right\rangle-\int_{-\infty}^0 p_0 \ d z+ \sigma \left\langle1-\cos\theta\right\rangle,
\end{equation}
where $\theta\edit{(x,t)}$ is the angle relative to the horizontal that the interface makes on the right (wavy) side of the control volume.  The pressure $p_o$ on the left side of the domain is simply hydrostatic, $p_o\edit{(z)}=-\rho g z$.  By considering small angles \edit{($1-\cos\theta\approx\frac{1}{2}\theta^2$ and $\theta\approx\frac{\partial\eta}{\partial x}$)}, and neglecting contributions of $O(A^3)$ henceforth, this can be rewritten as
\begin{equation}
S_{x x}= {\int_{-\infty}^0\left(\left\langle{p}+\rho{u^2}\right\rangle-p_0\right) d z}+\left\langle\int_{0}^{\eta}p \ dz\right\rangle+\frac{1}{2} \sigma \left\langle\left(\frac{\partial \eta}{\partial x}\right)^2\right\rangle. \label{eqn:Sxx}
\end{equation}
We compute the pressure $p\edit{(x,z,t)}$ in the wavy side by considering the vertical momentum balance on a vertical fluid segment,
\begin{equation}
p=-\sigma\frac{\partial^2 \eta}{\partial x^2}+\rho g(\eta-z)-\rho w^2. \label{eqn:pressure}
\end{equation}
In essence, the weight of the fluid segment must be balanced by the net pressure and vertical momentum flux. Averaging equation \ref{eqn:pressure} over a cycle yields
\begin{equation}
\left\langle p \right\rangle=-\rho g z-\rho \left\langle w^2 \right\rangle=p_o-\rho \left\langle w^2 \right\rangle, \label{eqn:pressureaverage}
\end{equation}
\edit{which is function of the depth $z$ only}. Substituting equations \ref{eqn:pressure} and \ref{eqn:pressureaverage} into equation \ref{eqn:Sxx}, and noting that $\left\langle u^2 \right\rangle =\left\langle w^2 \right\rangle$ for deep water, this reduces to 
\begin{equation}
S_{xx}=\frac{1}{2}\rho g \left\langle\eta^2\right\rangle-\sigma\left\langle\eta\frac{\partial^2 \eta}{\partial x^2}\right\rangle+\frac{1}{2} \sigma \left\langle\left(\frac{\partial \eta}{\partial x}\right)^2\right\rangle.
\end{equation}
The first two terms of this expression arise from the excess pressure term (second term in equation \ref{eqn:Sxx}) and represent contributions from hydrostatic and curvature pressures, respectively.  The final term comes from the oscillating line tension associated with the wavy interface.    Substituting in our expression for $\eta$ and averaging yields the final expression
\begin{equation}
    S_{xx}=\left(\frac{1}{4}\rho g+\frac{3}{4}\sigma k^2\right) A^2.\label{eqn:finalSxx}
\end{equation}
  \edit{The result for $S_{xx}$ is independent of $x$, otherwise momentum would accumulate in certain parts of the wave \cite{longuet1964radiation}.} To keep the raft in place due to the unbalanced momentum flux computed here, a force $S_{xx}$ must be applied to the raft in the positive $x$ direction to resist its tendency to propel in the $-x$ direction if unconstrained.  Note that due to the 2D nature of the problem considered here (as in \cite{longuet1964radiation}), $S_{xx}$ has units of force per unit length (i.e. depth into the page).  For finite fluid depth $H$, the first term in equation \ref{eqn:Sxx} does not vanish (as $\left\langle u^2 \right\rangle > \left\langle w^2\right\rangle$) and leads to a modified final result that is greater in magnitude than the deep water result (equation \ref{eqn:finalSxx}), but converges to it as $kH\rightarrow\infty$ \cite{longuet1964radiation}.  \edit{While we have assumed a constant amplitude propagating wave in 2D for the analysis in this section, in practice, the wave amplitude will decay away from an isolated wavemaker due to radial spreading (3D effects) and viscous attenuation.  Such effects are not accounted for in the theory of Longuet-Higgins and Stewart and would be interesting to consider in future modeling work.}

\subsection{Hydrodynamic Drag at the Interface}\label{sec:drag}

Regardless of its means of propulsion, when an object moves through a fluid at a speed $U$ it experiences hydrodynamic drag opposing its motion.  At a fluid interface this drag can manifest as subtle modifications to the many familiar forms associated with single-phase flows; however, new drag sources are also possible.  Relevant drag mechanisms are frequently delineated by an appropriately defined Reynolds number, with the geometry of the object itself also playing a deciding role in determining the principal resistance mechanisms.  As in \cite{ho2023capillary}, we define the Reynolds number as $Re=UL/\nu$, where $L$ is the length of the object in its direction of motion.

\subsubsection{Stokes' Drag}
At very low $Re$, interfacial particles experience Stokes' drag that is proportional to their velocity.  Considerable work has been done for the case of a spherical particle straddling an interface, where a modified Stokes drag law is frequently used
\begin{equation}
    F_{\mu}=\alpha 6 \pi \mu R U
\end{equation}
where the $\alpha$ parameter accounts for the presence of the interface, and can be either less or greater than the corresponding single-phase result ($\alpha=1$) depending on the parameters \cite{petkov1995measurement,danov2000viscous,zhou2022drag}.  Curiously, even though the particle may only be partially exposed to the more viscous fluid, the drag can be enhanced due to the flow associated with the highly curved meniscus \cite{petkov1995measurement}. This drag law has been applied in various models involving millimetric spheres moving at an air-water interface, and in practice, $\alpha$ is often determined indirectly from particle velocity measurements \cite{petkov1995measurement,vella2005cheerios,lagubeau2016statics,thomson2023nonequilibrium}.  The coefficient can also be predicted from theory and numerics \cite{danov2000viscous}, with the state-of-the-art summarized in the recent work by Zhou et al. \cite{zhou2022drag}.

\subsubsection{Form Drag}
At higher $Re$, and when a significant amount of the body is immersed in the water phase, pressure forces associated with flow separation (form drag) can be dominant.  The corresponding drag law takes the form 
\begin{equation}
    F_{D}=\frac{1}{2}C_D A \rho U^2 \label{eqn:FD}
\end{equation}
where $A$ is a reference area (often taken as the projected area in the streamwise direction), and $C_D$ is the drag coefficient.  Much is known about how $C_D$ depends on $Re$, body geometry, and surface roughness for single-phase flows, with far less known for interfacial bodies at our scales of interest. Recent experiments on spheres moving along an interface have shown the suitability of equation \ref{eqn:FD}, however the $C_D$ must also be modified to account for the object's degree of immersion, and the altered hydrodynamics associated with presence of the interface \cite{benusiglio2015wave,kamoliddinov2021hydrodynamic,hunt2023drag}.  Recent measurements by Hunt et al. \cite{hunt2023drag} for centimetric spheres at $Re=O(10^3)$ demonstrated that the drag coefficient can be more than three times that of the otherwise equivalent fully immersed case, with the surface wettability also playing an important role.  For the case of water-walking insects, form drag resistance is most commonly assumed in modeling efforts \cite{suter1997locomotion, bush2006walking,voise2010management}.

\subsubsection{Skin Friction}
For highly elongated bodies with shallow immersion, skin friction can play a dominant role.  For $Re\lesssim 10^6$ a laminar boundary layer forms along the surface of the body, growing in thickness roughly as $\delta \sim x^{1/2}$, where $x$ is the distance downstream from the leading edge.  Blasius' solution to the laminar boundary layer problem (2D) provides a drag law of form $f_{\delta}=0.664 \rho U^2 L / \sqrt{Re}$ \cite{kundu2024fluid}, yielding a power-law scaling of $U^{3/2}$ situated between the Stokes' drag ($\sim U$) and form drag ($\sim U^2$) cases discussed previously.  For a relatively flat object sitting atop a fluid interface (as in the case of a capillary disk), this shear frictional effect is the dominant resistance \cite{pucci2019friction}.  For the case of a circular disk ($L=2R$) Blasius' result can be integrated over the contact area yielding the prediction 
\begin{equation}
    F_{\delta}=1.64 \rho \sqrt{\nu} R^{3 / 2} U^{3 / 2},\label{eqn:blasius}
\end{equation}
 found to be in good agreement with velocity measurements of free capillary disks sliding along an air-water interface \cite{pucci2019friction}.  For non-circular objects, equation \ref{eqn:blasius} can be readily updated by taking both the geometry and its orientation into account \cite{pucci2019friction}.  This model has been used in rationalizing the interfacial motion of whirligig beetles \cite{jami2021overcoming} and wave-propelled rafts \cite{benham2024wave}. \edit{For intermediate Reynolds numbers ($1<Re<10^3$), additional corrections to Blasius' result due to the leading and trailing edges are anticipated to become relevant \cite{white2006viscous}.}

However, the fluid depth $H$ may be sufficiently thin such that the boundary layer (of maximum thickness $\delta_m \approx 5 \sqrt{\nu L/U}$) interacts with the bottom boundary.  In this case (i.e. when $\delta_m \gg H$) a linear drag law is more appropriate, resulting from the locally fully developed Couette-like flow formed between the underside of the body and the base of the container \cite{oza2023theoretical}.  This force can be approximated as
\begin{equation}
    F_H=\frac{\mu A}{H}U
\end{equation}
where $A$ is the area in contact with the fluid.  For the case of a circular capillary disk rotating at the interface, this shear effect introduces a resistant torque given by
\begin{equation}
    \tau_H=\frac{\pi\mu R^4}{2H}\Omega
\end{equation}
where $\Omega$ is the angular rotation rate of the disk.

\subsubsection{Wave Drag}
Objects moving along an air-water interface at a constant speed $U$ tend to generate capillary and gravity waves due to their motion.  Generation of surface waves costs energy, and this expense represents another form of interfacial hydrodynamic resistance referred to as wave drag \cite{havelock1917some}.  Unlike the other forms of hydrodynamic drag discussed thus far, wave drag is non-monotonic with speed, and generally peaks at an intermediate speed.  For large-scale objects this drag is generally characterized by the Froude number $Fr=\frac{U}{\sqrt{gL}}$, however for capillary-scale objects moving at speeds on the order of centimeters per second the effect can be more nuanced.  In particular, potential flow theory predicts that for speeds $U<c_{m}$, no waves will be excited and thus there is no contribution to wave drag \cite{raphael1996capillary}.  A number of measurements of this drag have been performed at small scales \cite{browaeys2001capillary,burghelea2001onset,burghelea2002wave,le2011wave} demonstrating the non-monotonic behavior and large increase as speeds exceed $c_m$.  Although a simple formula for wave drag does not exist, if one assumes a form of a pressure distribution moving along the interface at speed $U$, the wave resistance can be computed using Fourier transform methods \cite{raphael1996capillary,benzaquen2011wave}.  Nevertheless, whether wave drag is dominant in practice depends on the system parameters, geometry, and scale \cite{jami2021overcoming,hunt2023drag}.  The capillary-scale interfacial propulsors discussed here move well below $c_m$, and thus its potential influence has been neglected thus far in the literature.


\subsection{Interfacial Interactions}\label{sec:cheerios}

When multiple objects share a common fluid interface, lateral forces can be induced through the deformation field created by their presence.  At the capillary scale, identical axisymmetric objects create like-signed menisci that cause the particles to attract one another in an effort to minimize the total potential energy (gravitational and surface energies) of the system.  This effect is now commonly referred to as the ``Cheerios effect'' \cite{vella2005cheerios}, but has been a subject of scientific inquiry for almost a century. The phenomenon was leveraged by Bragg, who used it to assemble rafts of small monodisperse bubbles at a fluid interface demonstrating analogies with microscopic crystal physics \cite{bragg1942model,bragg1947dynamical}.  Inspired by this work, the capillary attraction effect was then rationalized mathematically by Nicolson in 1949, who demonstrated striking quantitative similarities between the interaction potential between millimetric bubbles and atomic interaction potentials defined by van der Waals attraction and short-range repulsion \cite{nicolson1949interaction}.  There is also evidence that insects at the air-water interface use capillary attraction to self-assemble \cite{voise2011capillary,loudet2011mosquito,ko2022small} and climb static menisci \cite{hu2005meniscus}.

For objects in static (or quasi-static) equilibrium, the capillary interaction length scale is defined by the capillary length $l_c$, consistent with the discussion in section \ref{sec:equilibrium}.  For an isolated axisymmetric particle, the height profile surrounding the object is given as the solution to the linearized Young-Laplace equation
\begin{equation}
    h(r)=Q\mathrm{K}_0(r/l_c)
\end{equation}
where the characteristic length $Q$ is determined by the vertical equilibrium condition, and is referred to as a capillary ``charge'' in analogy with electrostatics \cite{kralchevsky2001particles,delens2023induced}.  For the case of a circular capillary disk, $Q=-d/\mathrm{K}_0(R/l_c)$ by comparison to equation \ref{eqn:hr}, where $d$ is set by the vertical force balance in equation \ref{eqn:equilibrium2}.  An analogous expression for $Q$ in the case of a spherical particle can be found in prior works \cite{vella2005cheerios,delens2023induced}.  To leading order in $Bo$ (i.e. objects that are much smaller than the capillary length), and assuming the menisci of the two particles simply superimpose (referred to as the ``Nicolson approximation''), the potential energy of two identical objects placed a center-to-center distance $l$ apart is simply the effective weight (actual weight minus Archimedes force) of one particle multiplied by its vertical displacement due to the meniscus of the other particle \cite{vella2005cheerios}:
\begin{equation}
    E(l)=-2 \pi \sigma Q^2 \mathrm{K}_0\left(\frac{l}{l_c}\right).
\end{equation}
The lateral attraction force (capillary attraction) is then the gradient of this quantity
\begin{equation}
    F(l) =-\frac{dE}{dl}=-2 \pi \sigma \frac{ Q^2}{l_c} \mathrm{K}_1\left(\frac{l}{l_c}\right)\label{eqn:cheerios}
\end{equation}
where the overall negative sign indicates attraction.  Mathematically, when $l\ll l_c$, $\mathrm{K}_1\left(\frac{l}{l_c}\right) \approx \frac{l_c}{l}$, and equation \ref{eqn:cheerios} becomes analogous to a 2D version of the electrostatic interaction between point charges, hence the term capillary ``charges'' \cite{kralchevsky2001particles}.  A more general (vector) form of equation \ref{eqn:cheerios} is given by 
\begin{equation}
\mathbf{F}=F_p \nabla h\rvert_{\mathbf{x}=\mathbf{x_p}}\label{eqn:generalcheerios}
\end{equation}
where $F_p$ is the net vertical force on the particle, and $\nabla h\rvert_{\mathbf{x}=\mathbf{x_p}}$ is the local gradient of the interface induced by the presence of the other particle(s) \cite{dominguez2008force,de2018capillary}.  Such surface gradient-driven forcing is analogous to that typically assumed in models of walking droplets \cite{molavcek2013drops,oza2013trajectory}.  When the particles become very close, the superposition approximation breaks down, and the particles (and/or their contact lines) tilt, increasing the net attractive force \cite{ho2019direct,delens2023induced}.

While this framework is highly effective for very small particles ($R\ll l_c$), less is known at larger scales, although asymptotic expressions exist in certain cases \cite{he2013capillary}.  For the case of two circular capillary disks, scaling arguments combined with direct experimental measurements and simulations suggest an approximate attraction law of form 
\begin{equation}
    F(l)=-2F_{\sigma}e^{-{\frac{l-2R}{l_c}}} \label{eqn:cheeriosdisk}
\end{equation}
where 
\begin{equation}
    F_{\sigma}=\frac{(m g)^2 R^{1 / 2}}{\pi^2 \sigma l_c^{3 / 2}\left[\left(R / l_c\right)^2+2 R / l_c\right]^2}
\end{equation}
is the characteristic force scale \edit{(derived assuming $R\gg l_c$)}, and the factor of 2 in equation \ref{eqn:cheeriosdisk} is an empirical coefficient determined by fit to experiments and numerical simulations over a range of parameters (Figure \ref{fig:Fundamentals}(d)) \cite{ho2019direct}.  While not an exact result, this expression is a convenient and accurate formula for the range of $Bo\gtrsim1$ \cite{hooshanginejad2024interactions}.

Nevertheless, in all equilibrium cases discussed here, the capillary interaction length scale is set by $l_c$ and the attraction force is strictly attractive (note that two particles with opposite capillary charges will repel, but we have restricted our discussion to identical particles).  However, De Corato and Garbin considered the case where a particle is periodically forced normal to the interface, with applied force $F_p(t)=F_0\cos \omega t$ \cite{de2018capillary}.  By solving the potential flow equations for the wavy interface shape in steady state and applying equation \ref{eqn:generalcheerios}, they demonstrated that two driven particles could experience both cycle-averaged attraction and repulsion force due to the wavy interface, alternating on a characteristic lengthscale given by the capillary wavelength $\lambda$.  However, due to the application of a reflecting boundary condition at infinity (rather than a radiation boundary condition), they predicted a standing wavefield surrounding the particle.  This problem was later revisited with a Sommerfeld radiation condition, predicting an outwardly propagating wave as observed in the surfer experiments \cite{ho2023capillary}, and also updated to include the effects of gravity and weak viscosity \cite{oza2023theoretical}.  This follow-up work is reviewed in section \ref{sec:interaction}, and forms the current basis for modeling the wave interaction between capillary surfers and spinners. 

\section{Wave propulsion}\label{sec:propulsion}
\begin{figure}
    \centering
    \includegraphics[width=\linewidth]{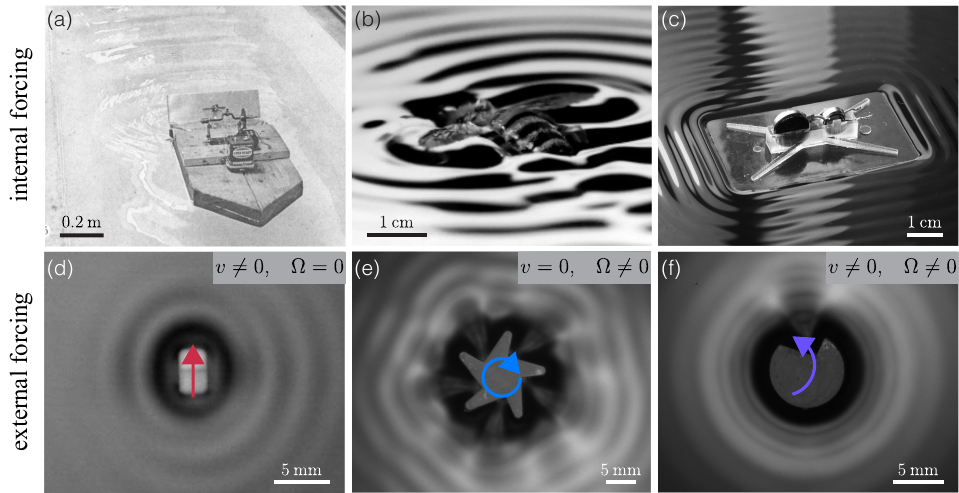}
    \caption{Wave propulsion resulting from unbalanced wave radiation stress excited by internal forcing (top row) and external forcing (bottom row) of floating objects. (a) A floating wavemaker self-propels in the direction opposite to wave propagation due to the net wave radiation stress (image reproduced from \cite{longuet1977mean}). (b) A honeybee pinned to a fluid interface generates a fore-aft asymmetric wavefield by flapping its wings, leading to thrust (image reproduced from \cite{roh2019honeybees}). (c) A small robotic device (``SurferBot'') consists of a rigid rectangular platform driven by a vibration motor and self-propels along a fluid interface \cite{rhee2022surferbot}. (d) An asymmetrically weighted capillary disk (capillary surfer) self-propels on a vibrating fluid bath \cite{ho2023capillary}. (e) A chiral star-shaped capillary disk (capillary spinner) rotates steadily in place on a vibrating fluid bath \cite{barotta2023bidirectional}. (e) Disks with both polar asymmetry and chirality propel on a vibrating fluid bath along curved arcs \cite{barotta2023bidirectional}.}
    \label{fig:propulsion}
\end{figure}

As reviewed in section \ref{sec:radiation}, propagating waves carry momentum with them and thus will impart a reaction force on a localized source of such waves.  In fact, the ejection of momentum underlies the physics of inertial propulsion in general.  Longuet-Higgins cleverly noted that ``the radiation stress may be actually set up to use for propelling a small craft'' and demonstrated the principle in experiment (Figure \ref{fig:propulsion}(a), \cite{longuet1977mean}).  Their ``wave-powered craft'' consisted of a planar wavemaker (driven by a crank mechanism at 3 Hz) mounted atop a freely floating wooden plank, and moved at speeds of 10-15 cm/s \edit{($\sim 0.2$ body lengths per second)} in the direction opposite of wave propagation.  A similar propulsion mechanism was later documented for a water-bound honeybee (Figure \ref{fig:propulsion}(b)), whose oscillating wings generated a fore-aft asymmetric wavefield \cite{roh2019honeybees}.  For this biological system, the frequencies observed were between 40 and 290 Hz, confidently in the capillary wave regime, with net translation speeds ranging from 1.4 to 4.3 cm/s \edit{($1-4$ body lengths per second)}.  Coherent steady streaming flows were documented in tandem with the waves, but represented a smaller contribution to the overall thrust than wave radiation (estimated using equation \ref{eqn:finalSxx}).  While capillary wave generation has been known to play some role in the interfacial propulsion of other interface-dwelling insects, it is generally subdominant to other mechanisms such as vorticity generation \cite{hu2003hydrodynamics, bush2006walking, buhler2007impulsive} (although the ordering can reverse for 2D idealizations \cite{gao2011numerical}).  The partitioning of momentum between waves and vortices remains an active topic of discussion \cite{steinmann2018unsteady}.

More recently, a simple robotic device demonstrated the possibility of wave propulsion in the case of rigid-body motion of a floating object (Figure \ref{fig:propulsion}(c), \cite{rhee2022surferbot}).  This device, referred to as the ``SurferBot,'' consists of a small vibration motor fixed near one end of a centimeter-scale 3D-printed floating plate. Direct experimental measurements confirmed the generation of a fore-aft asymmetric capillary wavefield by the bobbing and pitching plate.  Vibration motors were selected with a frequency around 80 Hz, and resulted in mean propulsion speeds of 1.8 cm/s \edit{($0.4$ body lengths per second)}.  The wave-propulsion force was estimated using the experimental wavefield measurements and equation \ref{eqn:finalSxx} to be around 8 dynes (80 $\mu$N), of similar magnitude to analogous estimates provided for the interface-bound honeybee \cite{roh2019honeybees}.  Around the same time, wave propulsion at human scales was documented for the case of ``gunwale bobbing,'' similarly achieved by vertically forcing a floating craft towards one end \edit{with cruising speeds around 1 m/s ($\sim 0.2$ body lengths per second)} \cite{benham2022gunwale}.  Rather than appealing to momentum flux arguments, the propulsion was rationalized using the wave drag theory of Havelock \cite{havelock1919wave}, modeling the craft as an oscillatory pressure forcing at the interface.  \edit{In this case, the speed of the propulsor is faster than the wave group velocity (supercritical) and thus the wavefield takes a distinctly different form than that of the Surferbot, which moves notably slower than the group velocity (subcritical) \cite{benham2024wave}.  As later discussed in section \ref{sec:efficiency}, the non-dimensional Mach number ($Ma$) delineates these regimes and is anticipated to be closely related to the propulsion efficiency \cite{benham2024wave,o2025achieving}.}

For the systems discussed thus far, the particle is driven {\it internally} with an onboard actuator and energy source.  The wave-propulsion mechanism can also be realized by oscillating the environment, allowing for wave propulsion of significantly smaller particles.  In the work of Ho et al., asymmetric millimeter-scale capillary disks were placed atop a vibrating fluid bath and shown to self-propel with speeds up to 3 mm/s \edit{($0.7$ body lengths per second)} (Figure \ref{fig:propulsion}(d), \cite{ho2023capillary}).  The disks were designed to be heavier on one end, and thus the asymmetric inertial body forcing led to asymmetric rigid-body kinematics. Although direct wavefield measurements were not conducted as part of this initial study, the measured parametric dependence of the propulsion speed was shown to be consistent with an underlying wave-propulsion mechanism (section \ref{sec:radiation}) resisted by a linear viscous drag along the base of the surfer (section \ref{sec:drag}).  Immiscible droplets floating on a vibrating viscous bath have also been shown to self-propel at the interface due to radiation stress, however, in this scenario the waves are standing waves on the interface of the propelling droplet itself \cite{pucci2015faraday}.

Fully predictive models of the coupled fluid-structure dynamics associated with wave-driven propulsion are scarce at the moment, although a first effort has recently been put forth by Benham et al. \cite{benham2024wave,benham2025wave}.  In this work, they numerically modeled the coupled raft and fluid dynamics in 2D using a quasi-potential free surface model (discussed in section \ref{sec:wavegravcap}) coupled to rigid-body dynamics for the raft.  Predictions for the net thrust were then balanced with a skin friction drag law (discussed in section \ref{sec:drag}).  Using parameters provided in the experimental work, they modeled the wave-propulsion of the SurferBot using their framework and predicted propulsion speeds similar to those observed in experiment.  They also demonstrated consistency between their mean thrust force predictions (obtained by integrating the projected hydrodynamic pressure on the solid body) and the wave momentum theory of Longuet-Higgins \cite{longuet1964radiation} reviewed in section \ref{sec:radiation}.

It has thus been clearly established in a range of systems \edit{spanning millimeter to meter scales} that an asymmetric wavefield can enable steady linear propulsion, \edit{with speeds typically on the order of one body length per second or less}.  Extending this growing body of work, Barotta et al. demonstrated the possibility of achieving steady {\it rotation} using net torques arising from unbalanced wave-radiation stress (Figure \ref{fig:propulsion}(e), \cite{barotta2023bidirectional}).  In this work, chiral star-shaped capillary disks (capillary spinners) were placed atop a vibrating bath, and shown to rotate in place at a constant rate.  While the objects rotated monotonically faster with the driving amplitude, their rotation direction was shown to depend on both the size of the spinner and the frequency (with the latter setting the wavelength, as reviewed in section \ref{sec:wavegravcap}).  By appealing to a simplified model of the self-excited wavefield, it was demonstrated that the interplay between the geometry of the particle and its self-excited wavefield dictate the net wave torque.  In particular, constructive and destructive wave interference endows the wavefield with a finescale frequency-dependent structure responsible for the bidirectional motion.  The effects of chirality and polar asymmetry were then combined to realize a particle that moves along curved trajectories, with the turning direction and radius of curvature remotely controllable by the driving frequency (Figure \ref{fig:propulsion}(f),\cite{barotta2023bidirectional}).  Sustained rotary motion was also documented in more recent work by Sungar et al. \cite{sungar2025synchronization}, where the spinners were achiral in plan view however each arm consisted of a slanted wedge in contact with the underlying fluid.  While these latter externally driven systems have been achieved by mechanical vibration, analogous wave propulsion can be realized by applying an AC magnetic field to floating asymmetric magnetic particles \cite{metzmacher2023assembly}.

\section{Wave interaction}\label{sec:interaction}

\begin{figure}
    \centering
\includegraphics[width=\linewidth]{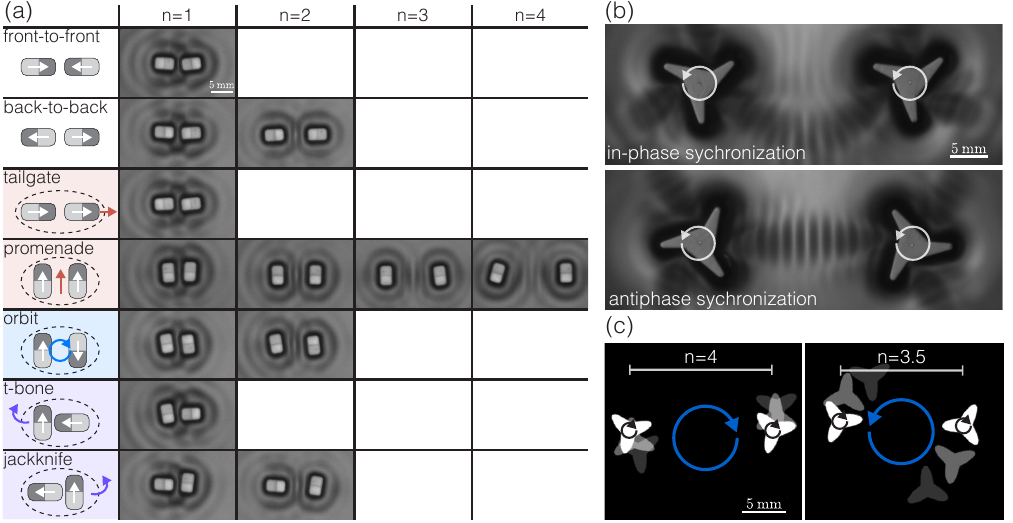}
    \caption{Wave-mediated interaction between pairs of identical surfers and spinners. (a) Pairs of surfers exhibit seven qualitatively distinct stable interaction modes for a fixed set of surfer, fluid, and driving parameters.  Each pairwise mode can occur at multiple discrete spacings, separated by approximately an integer number of wavelengths, $\sim n\lambda$ \cite{ho2023capillary,oza2023theoretical}.  (b) Capillary spinners magnetically tethered at a fixed separation synchronize their rotation in either an in-phase or antiphase configuration depending on their initial conditions \cite{barotta2025synchronization}.  (c) Free capillary spinners find quantized lateral spacings while simultaneously synchronizing their individual rotations, and also executing a comparatively slow global precession \cite{sungar2025synchronization}.}
    \label{fig:waveinteraction}
\end{figure}

As outlined in the previous section, objects at a fluid interface can steadily propel in response to self-generated outwardly propagating waves.  These waves can extend over an impressive range along the interface, communicating dynamic stresses to other interfacially collocated bodies.  In the world of surface-dwellings insects, surface waves are used for detection of both prey and predators, as well as for communication \cite{wilcox1972communication,lang1980surface,bush2006walking,voise2014echolocation}.  Wave-emitting vibrating robots can also be attracted or repelled to nearby boundaries due to wave-interaction effects \cite{tarr2024probing}.

Interactions between wave-driven particles has been studied for the externally driven capillary surfers and spinners.  When two surfers approach one another they may either scatter or lock into one of many stable equilibrium configurations (Figure \ref{fig:waveinteraction}(a), \cite{ho2023capillary}).  Seven qualitatively different interaction modes were documented for the surfer geometry considered in this work, with the stability of the modes depending on the system parameters.  While a single surfer in isolation moves along a straight linear trajectory, pairs of surfers can also become static, rotate about their center of mass, or move along arced trajectories, depending on their configuration.  Collective interactions thus broaden the possible propulsion modalities of a single surfer in isolation.  Furthermore, each qualitative mode can exist with multiple stable spacings, where the nearest faces are roughly parallel and separated by approximately an integer number of wavelengths, denoted by $n$.  While similar quantized spacings have been documented for the case of walking droplets \cite{couder2005walking,protiere2006particle}, such droplets principally interact through a standing wavefield \cite{eddi2011information}, whereas here each particle emits a strictly propagating wavefield.  As elucidated in the modeling efforts presented later in the section, the combination of the synchronized vertical oscillation of the particle and the incident periodic waves results in a non-zero time-averaged lateral wave force, with stable equilibria separated by integer numbers of wavelengths.  Larger assemblies were also documented experimentally with up to eight surfers moving collectively.  While such spatially periodic pairwise interactions might appear peculiar to wave-mediated systems, periodic vortex wakes associated with flapping flight and fish swimming also lead to similar quantized spatial order and emergent pattern formation \cite{ramananarivo2016flow,ormonde2024two}.

Pairs of capillary spinners magnetically tethered at a fixed distance apart were explored in later work by Barotta et al. (Figure \ref{fig:waveinteraction}(b), \cite{barotta2025synchronization}).  The wave interaction directly leads to rotational phase synchronization, which at certain inter-spinner spacings is bistable with both in-phase and antiphase synchronization possible (depending on the initial conditions).  Furthermore, non-identical spinners can also synchronize provided their intrinsic differences are not too disparate, in analogy with the classical Kuramoto phase oscillator \cite{acebron2005kuramoto}.  If the wave coupling becomes too strong, the spinners can lock each other in place, stifling rotation entirely.  The key behaviors documented in this work again occurred with a distinct spatial periodicity, with the period set by the wavelength.  Pairs of untethered spinners were considered by Sungar et al. \cite{sungar2025synchronization}, and shown to adapt the documented behaviors of both the free surfers and tethered spinners: they find stable quantized spacings based on the wavelength, while also synchronizing in both in-phase and antiphase configurations (Figure \ref{fig:waveinteraction}(c)).  Similar behaviors were noted for both co-chiral and anti-chiral pairs. The authors also made connection to ``swarmalators'' as a system that simultaneously swarms and synchronizes through coupled spatial and phase dynamics \cite{o2017oscillators}. Both integer and half-integer spacings are documented in this system, and corresponded to in-phase and antiphase synchronization, respectively. Curiously the pairs also exhibit a slow global rotation, with such drift documented in the same direction as an individual spinner's rotation for in-phase spinners and in the opposite direction for antiphase spinners.  Larger collections of free spinners (up to four) were documented to have similar behaviors to pairs: quantized spacing, phase synchronization, and slow global rotation.

\subsubsection{Wave-Interaction Modeling}

\begin{figure}
    \centering
\includegraphics[width=\linewidth]{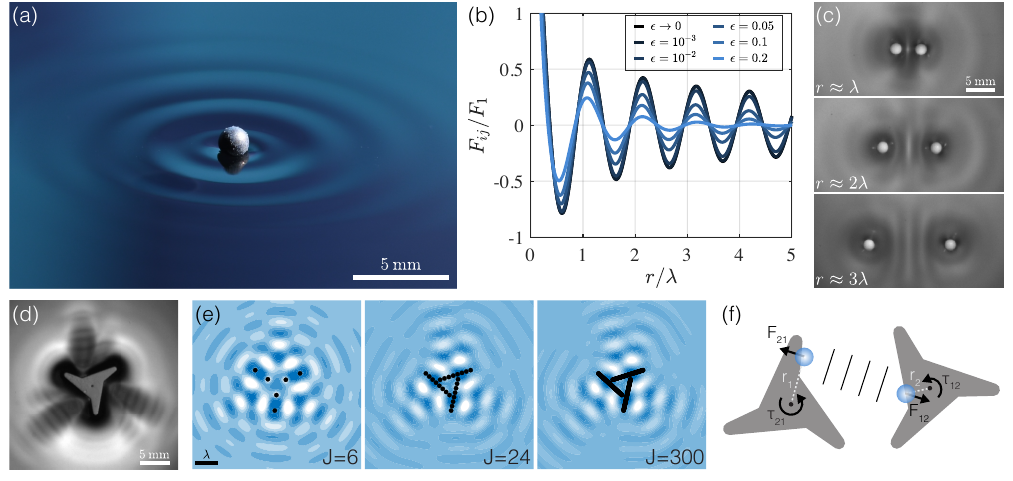}
    \caption{(a) A single spherical particle on a vibrating fluid bath ($f=100$ Hz) emits an extended capillary wavefield. (b) The predicted wave-induced interaction force between two periodically driven point-sources for different magnitudes of non-dimensional viscosity $\epsilon$ (equation \ref{eqn:force}, $\beta=0.048$) \cite{oza2023theoretical}. (c) The capillary attraction between two particles on a vibrating bath is stabilized by wave effects, with stable equilibrium spacings quantized by the wavelength $\lambda$. \edit{(d) The wavefield of a three-armed capillary spinner at $f=90$ Hz \cite{barotta2025synchronization}. (e) The wavefield of a simulated capillary spinner comprised of $J=6,24,300$ point sources, respectively. One wavelength is indicated by the black scale bar. (f) A pair of interacting spinners interact via pairwise interactions of their constituent point sources yielding both resultant forces and torques on the bodies.} }
    \label{fig:waveinteractionmodel}
\end{figure}

Before modeling interactions between wave-exciting particles or propulsors, we first consider the waves generated by a single periodically driven point source.  This problem was considered mathematically by De Corato and Garbin \cite{de2018capillary} with the extension by Oza et al. \cite{oza2023theoretical} reviewed here.  The governing equations are as follows, where the linearized fluid flow can be expressed in term of the velocity potential $\phi$ ($\mathbf{u} = \nabla \phi)$ and wave height $h$:
\begin{align}
\Delta\phi + \phi_{zz} &= 0, \quad \text{for} \quad z<0, \quad \mathbf{x} \in \mathbb{R}^2\\
\phi_t  &= \frac{\sigma}{\rho}\Delta h + 2\nu\Delta\phi -gh+  \frac{F_0}{\rho}\cos\omega t \ \delta(\mathbf{x}), \quad \text{at} \quad z=0 \label{eqn:dynamicBC}\\
    h_t  &= \phi_z + 2\nu \Delta h, \quad \text{at} \quad z=0\\
    \phi  &\to 0, \quad h\to 0 \quad \text{as} \quad |\mathbf{x}|,-z\to \infty.
\end{align}
Here, $\Delta = \partial_{xx}+\partial_{yy}$ and the boundary conditions at the interface are evaluated at the mean interfacial height of $z=0$.  The localized point forcing is represented as the final term in the dynamic boundary condition (equation \ref{eqn:dynamicBC}).  The model includes viscous corrections that enter into the dynamic and kinematic boundary conditions as derived from the quasi-potential theory (discussed in section \ref{sec:wavegravcap}), incorporating the leading order effect of viscosity concentrated in the boundary layer at the oscillating interface.  

From here, the steady-state wavefield can be solved for and expressed as $h(r,t) = \text{Re}\left[ H(r)e^{i\omega t}\right]$ with
\begin{equation}\label{eq:wavefield}
H(r) = \frac{F_0}{12\sigma} \sum_{p=1}^4 \frac{H_0(-k_pk_cr) - Y_0(-k_pk_cr)}{1+(4/3)i\epsilon/k_p + (4/3)\epsilon^2k_p+\beta/3k_p^2},
\end{equation}
and the $k_p$s are the roots to the fourth order polynomial equation $p(k;\epsilon,\beta) = \epsilon^2k^4 + 2ik^2 + k^3 +\beta k - 1$.  Here, $l_c= \sqrt{\frac{\sigma}{\rho g}}$ is the capillary length as previously defined and $ k_c = \left(\frac{\rho \omega^2}{\sigma}\right)^{1/3}$, which is the wavenumber in the capillary deep-water limit of the inviscid dispersion relation (equation \ref{eqn:dispersion}).  The surfer experiments take place in the deep-water capillary regime.  Disregarding the dimensional prefactor, the overall shape of the waveform is ultimately governed by two dimensionless numbers: $\epsilon  = \frac{2\nu k_c^2}{\omega}$ and $\beta =(k_cl_c)^{-2}$.  Physically, $\epsilon$ is a dimensionless viscosity or equivalently an inverse wave Reynolds number (as introduced in section \ref{sec:wavegravcap}). The second parameter $\beta$ is a wave Bond number, and compares the lengthscale associated with the wavefield ($k_c^{-1}=\lambda/2\pi$) to the capillary length.  The wavefield $h(r,t)$ takes the form of an outwardly propagating axisymmetric monochromatic wave.

For a small solid particle of mass $m$ at the fluid interface (Figure \ref{fig:waveinteractionmodel}(a)), the effective localized forcing amplitude is assumed to be $F_0=\alpha m \gamma$, where $\gamma$ is the acceleration amplitude of the bath. The non-dimensional prefactor $\alpha$ is of order one and accounts for the unresolved vertical dynamics of the particle, inherently coupled to the fluid bath.  In prior works, this factor has either been assumed to be identically one \cite{oza2023theoretical} or determined via fit to direct wavefield measurements \cite{barotta2025synchronization}.  With an expression for the wavefield available, the interaction force between two interacting point sources can be computed using equation \ref{eqn:generalcheerios}.  To this end, we consider two oscillating point sources located at $\mathbf{x}_i$ and $\mathbf{x}_j$, undergoing in-phase vertical dynamics according to $\ddot{z}_i=\ddot{z}_j=-\alpha \gamma \cos (\omega t)$.  We can then compute the time-averaged force on point source $i$ of mass $m_i$ due to the deformation of the interface by point source $j$ of mass $m_j$. Using the expression for the wavefield (equation \ref{eq:wavefield}), the cycle-averaged wave interaction force can then be computed as
 \begin{equation}
        \mathbf{F}_{\text{w},ij}(r) = \left\langle\left. m_i \ddot{z}_i \nabla h\left(\mathbf{x}-\mathbf{x}_j, t\right)\right|_{\mathbf{x}=\mathbf{x}_i}\right\rangle = F_1\sum_{p=1}^4 \text{Re}\left[ k_p
 \frac{H_{-1}(-k_pk_cr)+Y_{1}(-k_pk_cr)}{1+(4/3)i\epsilon/k_p + (4/3)\epsilon^2k_p+\beta/3k_p^2}\right]\hat{\mathbf{r}}_{ij},
 \label{eqn:force}
\end{equation}
where $F_1 = \alpha^2 m_im_j\gamma^2k_c/24\sigma $ and $\hat{\mathbf{r}}_{ij} = \frac{\mathbf{x}_j-\mathbf{x}_i}{|\mathbf{x}_j-\mathbf{x}_i|}$.  Here, a positive force corresponds to net attraction whereas a negative force corresponds to repulsion.  Note that if particle $i$ were not itself oscillating, the model would predict no average interaction force, as it would uniformly sample all gradients of the incident propagating wave.  However the synchronized oscillation results in a non-uniform dynamic sampling of the incident wavefield, and leads to the non-zero time average given by equation \ref{eqn:force}.  Sample interaction force predictions are provided in Figure \ref{fig:waveinteractionmodel}(b).  The force law predictions regions of both net attraction and repulsion, with the roots representing equilibria.  Stable equilibria occur at approximately integer multiples of the wavelength.  These quantized stable equilibria can be readily observed in experiment by placing small hydrophobic beads on the surface of a vibrating bath, as demonstrated in Figure \ref{fig:waveinteractionmodel}(c).  The wave interaction force must be sufficient to overcome the static capillary attraction, which for the case of two identical circular capillary disks is given by
 \begin{equation}
        \mathbf{F}_{\text{c},ij}(r) = \frac{2\pi \sigma d^2}{\mathrm{K}_0^2\left(\frac{R}{l_c}\right)}\mathrm{K}_1\left(\frac{r}{l_c}\right)\hat{\mathbf{r}}_{ij}
 \end{equation}
 where $d$ is defined by the equilibrium force balance (equation \ref{eqn:equilibrium2}).  This purely attractive capillary force decays over the capillary length $l_c$, and can be neglected when $r\gg l_c$ \cite{barotta2025synchronization}.

To extend the framework for the interaction of two point sources to the wave-propelled systems of interest here, we model each body as a finite collection of point sources spatially arranged and geometrically constrained in a way as to heuristically capture the actual geometry of the wave-emitting object (\edit{Figure \ref{fig:waveinteractionmodel}(d,e)}).  \edit{In particular, as the density of the point sources is increased, the predicted wavefield converges as the inter-source spacing becomes much smaller than the wavelength.} For the case of pairwise interactions of two bodies, the point sources on an object $n$ interact with the point sources composing the second object.  As established in the first work on modeling surfer interactions \cite{oza2023theoretical} and later applied in the works on tethered and free spinners \cite{barotta2023bidirectional,sungar2025synchronization}, we can write the governing ODEs for the translational ($\mathbf{x}=(x,y)$) and angular ($\theta$) degrees of freedom for two identical particles ($n=1,2$) of mass $m$ and moment of inertia $I$ as the following:
\begin{align}
         m\ddot{\mathbf{x}}_n +D\dot{\mathbf{x}}_n &= DU_n\hat{\mathbf{n}}_n+\sum_{i=1}^{J}\sum_{j=1}^{J}  \left(\mathbf{F}_{\text{w},ij}+\mathbf{F}_{\text{c},ij}\right), \label{eqn:translation} \\
I\ddot{\theta}_n +D_R\dot{\theta}_n  &= D_R\Omega_n+ \sum_{i=1}^{J}\sum_{j=1}^{J} \left[ \mathbf{r}_{i} \times  \left(\mathbf{F}_{\text{w},ij}+\mathbf{F}_{\text{c},ij}\right)\right] \cdot \hat{\mathbf{z}}. \label{eqn:rotation}
    \end{align}
For this model, an isolated object moves with a free speed $U_n$ along its orientation $\hat{\mathbf{n}}_n$ and rotates about its center of mass with angular velocity $\Omega_n$.  As such, propulsion forces and torques are imposed rather than predicted here, with the numerical values determined from independent speed measurements on an isolated object.  The sums represent the force and torque contributions from the $j=1...J$ point sources composing the second object acting on each of the $i=1...J$ point sources of the object $n$ itself, where $\mathbf{r}_{i}$ represents the position of point source $i$ relative to the center of mass \edit{(an example for the case of two spinners is shown schematically in Figure \ref{fig:waveinteractionmodel}(f))}.  The linear ($D$) and rotational ($D_R$) drag coefficients can be modeled by assuming a locally fully developed Couette flow beneath the object \cite{oza2023theoretical} (discussed in section \ref{sec:drag}), or measured using a free deceleration experiment of an isolated propulsor \cite{barotta2025synchronization}.  By balancing the viscous drag with inertia, a viscous timescale $t_\nu$ can be formed, which for the surfers and spinners studied thus far is approximately 0.5 seconds.  The model additionally assumes that the center of friction occurs at the center of mass, otherwise additional couplings may arise such as a self-aligning torque \cite{baconnier2025self}.

For the case of pairwise surfer interactions, each surfer was modeled as two unequal point sources and $\Omega_n=0$ \cite{oza2023theoretical}.  The model recovered all of the experimentally observed pairwise interaction modes as well as identified numerous unstable equilibrium configurations. All equilibrium states were analyzed using linear stability analysis.  For the case of tethered spinners, only equation \ref{eqn:rotation} (rotational dynamics) needed to be considered for each spinner \cite{barotta2023bidirectional}.  Each spinner was modeled as $J=24$ point sources, with the predicted wavefield showing good qualitative agreement with visualizations and wavefield reconstructions.  Simulations of the two coupled ODEs recovered the experimentally observed synchronization and locking behaviors, and featured a similar spatial periodicity.  The case of free spinners was modeled using both the translation and rotational equations again, representing each spinner as four point sources, and taking $U_n=0$ \cite{sungar2025synchronization}.  Simulations recovered the simultaneous spatial quantization and orientational synchronization.  While many of the key experimentally observed interaction behaviors have been captured by the model presented in equations \ref{eqn:translation} and \ref{eqn:rotation}, some effects are inaccurately predicted or notably absent.  For instance, the speeds of individual surfers in bound pairs can vary from their free speeds, and are not quantitatively aligned with the model predictions \cite{oza2023theoretical}.  This discrepancy is presumably due to the fact that the model of the wave propulsion is highly simplified.  The slow global precession observed for free spinners is also not captured by the model \cite{sungar2025synchronization}.  This limitation was attributed to higher order interaction effects not accounted for in the present model, such as induced surface flows or nonlinear wave effects.  Furthermore, additional work is needed to rigorously define a protocol for the placement and strength of sources when modeling an arbitrary wave-emitting geometry.

\section{Outlook and Future Directions}\label{sec:outlook}
In this section we outline a number of topics that have been touched upon or revealed in the described work but deserve future attention.  Such topics span across fluid mechanics, engineering, and fundamental physics, with many of the central concepts overlapping.  Each subsection is not intended to be an exhaustive literature review of the respective subject, but rather provide some initial insight while establishing potential connections to the surfer system.

\subsection{Surface Flows}

    \begin{figure}
        \centering
        \includegraphics{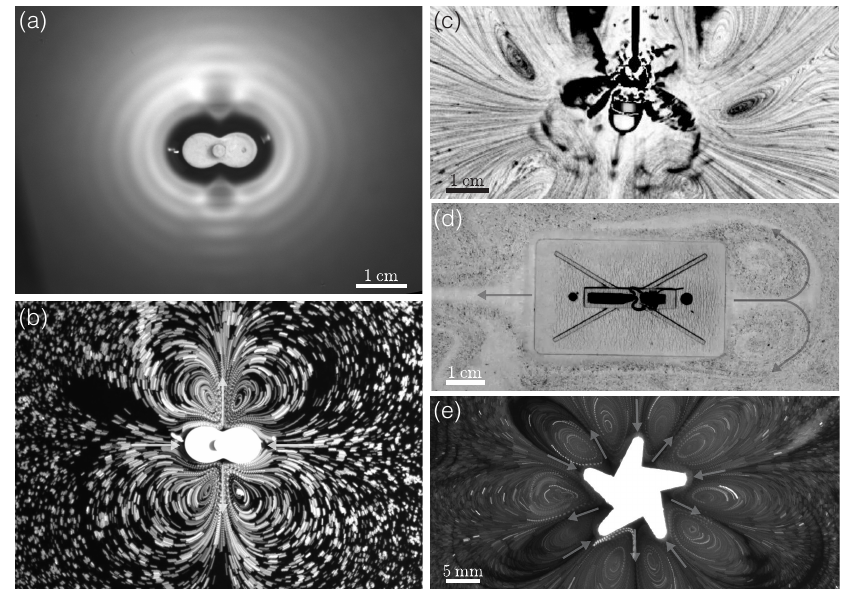}
        \caption{A peanut-shaped capillary disk on a vibrating bath generates (a) a complex wavefield that induces (b) strong surface flows. Similar wave-induced flows have been documented for the (c) trapped honeybee (image reproduced from \cite{roh2019honeybees}), (d) SurferBot \cite{rhee2022surferbot}, and (e) capillary spinners \cite{barotta2023bidirectional}.}
        \label{fig:flows}
    \end{figure}

    To this point, we have established that the emission of outwardly propagating capillary-gravity waves from an asymmetric object can lead to wave-driven propulsion and wave-mediated interactions. A byproduct of the wave generation is the formation of large-scale flows concentrated near the air-liquid interface.  A simple demonstration of this effect is shown in Figure \ref{fig:flows}(a,b).  A peanut-shaped capillary disk is placed atop a vibrating bath and generates a capillary wavefield as shown in Figure \ref{fig:flows}(a).  In particular, the concave portion focuses the generated waves into strong constructive beams along the central axis.  By placing small tracer particles at the interface, the streaklines associated with the concomitant surface flows are visualized in Figure \ref{fig:flows}(b). Pronounced outward jets are generated in the location of the wave beams, with the surface flows turning and eventually returning at the convex ends of the object.  In general, the overall structure of the flow pattern is dictated by the geometry of the object and its waves \cite{punzmann2014generation}.  Observations of large surface currents resulting from oscillating wavemakers have been documented in a variety of setups including for a cylinder \cite{punzmann2014generation}, parabola \cite{chavarria2018geometrical}, and orthogonal planar wavemakers \cite{francois2017wave}.  Of most relevance to the current work, robust surface flows were also documented for the case of the trapped honeybee \cite{roh2019honeybees}, SurferBot \cite{rhee2022surferbot}, and capillary spinners \cite{barotta2023bidirectional} (Figure \ref{fig:flows}(c-e)), as well as for small boats driven by AC-electrowetting \cite{chung2009electrowetting, yuan2015mechanism}.  These wave-induced flows have been shown to be important in other applications, such as mixing in agitated bioreactors with free surfaces \cite{bouvard2017mean, kim2025simulation} and the feeding mechanisms of freshwater snails \cite{joo2020freshwater}.
    
    Such streaming flows are inherently nonlinear wave effects and have been shown to arise due to a combination of Stokes drift and Eulerian mean flow contributions \cite{parfenyev2020large, abella2020measurement}.  While Stokes drift is perhaps the more familiar mechanism \cite{van2018stokes}, the Eulerian flows arise through a nonlinear generation of vertical vorticity \cite{filatov2016nonlinear}. The oscillating interface leads to a ``tilt'' in the vorticity concentrated within the thin viscous boundary layer at the surface yielding large-scale surface flows. Such flows are an example of steady streaming: the creation of a secondary flow with non-zero mean from an oscillatory source \cite{riley2001steady}. Although our focus here is at a free surface, steady streaming has been demonstrated as a propulsion mechanism in the bulk at intermediate Reynolds numbers  \cite{klotsa2015propulsion, dombrowski2020kinematics,derr2022reciprocal}.
    
    For both contributions, the characteristic speed of the surface flows is given by $u \sim \omega k A^2 =  c\left(kA\right)^2$, where $c$ is the phase velocity of the waves \cite{filatov2016nonlinear}. For the sake of arriving at some simple estimates, we focus on the Stokes drift contribution in the next portion, although Eulerian mean flows are very likely also significant in our systems of interest.  In general, the flow associated with Stokes drift is given as 
    \begin{equation}
\mathbf{u}_s = \left\langle \left(\int_0^t \mathbf{u} \ \text{d}t'\right) \cdot \mathbf{\nabla \mathbf{u}} \right\rangle,
\end{equation}
where $\mathbf{u}$ is the velocity field of the linearized wavefield \cite{van2018stokes}.  Although the Eulerian velocity at each point in space may be zero, a Lagrangian tracer that undergoes a finite excursion away from its initial position will sample different flow speeds during its trajectory if $\nabla \mathbf{u} \neq 0$, opening the possibility of a net cyclical drift.  For the case of a propagating plane wave, $\eta(x,t)=A\cos\left(kx-\omega t\right)$, the Stokes drift velocity is simply
\begin{equation}
\mathbf{u}_s = c\left(kA\right)^2e^{2kz} \ \hat{\mathbf{x}} \label{eqn:planeStokes}
\end{equation}
with the net flow moving in the direction of wave propagation $\hat{\mathbf{x}}$ \cite{kundu2024fluid}.  At the surface $(z=0)$ the flow has strength $c\left(kA\right)^2$ and decays into the bulk with a characteristic lengthscale $1/2k$. (We note that while the Eulerian contribution features the same characteristic velocity \cite{filatov2016nonlinear}, the penetration depth can depend on the spatial structure of the wavefield in addition to the wavenumber \cite{parfenyev2020large}).  The capillary waves generated in the wake of the SurferBot \cite{rhee2022surferbot} had a wavelength $\lambda = 0.40$ cm and amplitude $A = 0.02$ cm.  Using equation \ref{eqn:planeStokes}, and the capillary-wave dispersion relation, we can compute a characteristic Stokes drift velocity of $u_s = 3.2$ cm/s, which is of similar magnitude to the experimental flow measurements behind the surfer (and the speed of the SurferBot itself).

Although waves have been established as a dominant propulsion mechanism for the systems described in section \ref{sec:propulsion}, a natural follow-up question is whether such flows may also contribute a significant propulsion force.  For the case of the honeybee, the induced flows were measured and estimated to be a smaller contribution to the overall thrust than wave momentum, but of similar order of magnitude \cite{roh2019honeybees}.  Furthermore, in the AC-electrowetting propelled craft, the thrust was entirely attributed to the induced flows \cite{yuan2015mechanism} (although the possibility of wave momentum was not considered in the analysis).  Following the analysis put forth by Yuan et al. \cite{yuan2015mechanism} for the case of a plane wave, one finds a propulsive thrust (per unit depth into the page) associated with the Stokes drift momentum flux as 
\begin{equation}
\frac{F_{s}}{W} = \int_{-\infty}^0 \rho u_s^2 dz = \frac{\rho c^2 \left(kA\right)^4}{4k}.
\end{equation}
Given that the Stokes flow propulsion scales with $(kA)^4$ \cite{yuan2015mechanism}, and wave radiation stress propulsion scales with $(kA)^2$ \cite{pucci2015faraday, roh2019honeybees, ho2023capillary}, we anticipate that wave-driven propulsion to be the dominant mechanism in the small-amplitude limit.  
Nevertheless, we expect that as one increases the amplitude of the waves such that the small-amplitude assumption no longer is valid, nonlinear interactions that couple the flow and the wavefield could potentially lead to induced flows becoming the dominant mechanism for propulsion (see \cite{chavarria2018geometrical} for a detailed experimental investigation of wave-flow interactions with a parabolic wavemaker beyond the small-amplitude regime).

Lastly, while a theory for the nonlinear generation of vertical vorticity has been recently established \cite{filatov2016nonlinear}, the predictions have only been compared to experimental wavefields with relatively simple geometries.  Future experimental work featuring comparisons to theory should be completed for more complex wavemaker geometries to better understand their origin and relevance in the wave-propelled systems reviewed herein.  Furthermore, with an improved understanding of these flows, we can harness greater control over the wave and flow structure to manipulate passive objects along prescribed paths \cite{punzmann2014generation,wang2025topological} and enrich the concept of liquid-interface metamaterials \cite{francois2017wave}. 

\subsection{Wave-Structure Interaction}\label{sec:WSI}
The topic of wave-structure interaction (WSI) is one with notable historical attention due to its relevance to ships, offshore structures, and wave-energy harvesting.  Due to the scale of the typical applications involved, work has almost exclusively focused on large structures interacting with gravity waves, where surface tension has no consequence in either the fluid or structure dynamics.  Such interactions are frequently modeled using potential flow coupled to the rigid-body dynamics of a floating object \cite{linton2001mathematical}.  

In such analyses, it is commonly assumed the object undergoes small amplitude harmonic motion about its various degrees of freedom, with the computed dynamic effect of the surrounding fluid decomposed and subsequently interpreted in terms of an in-phase component (added mass) and a 90-degree out-of-phase component (damping, via radiated wave energy). Perhaps the simplest such case is that of an axisymmetric vertically heaving floating cylinder \cite{miles1987surface}.  Such an interpretation makes for convenient analogy with a simple harmonic oscillator, where the effect of the fluid is incorporated in the body dynamics as effective structural parameters.  For instance, one might model the vertical heaving motion of a rigid cylinder by the following ODE
\begin{equation}
    m_e\ddot{z}+C_e\dot{z}+k_ez = F_{ext}
\end{equation}
where the effective mass $m_e$ is the actual mass $m$ plus the added mass $m_a$, the damping coefficient $C_e$ accounts for the radiated wave energy, and the spring constant $k_e$ results from the vertically stabilizing effect of buoyancy and surface tension.  Although the WSI problem may itself be linear, the added mass and damping coefficients generally depend on frequency \cite{miles1987surface}.  As discussed throughout, analogous problems at smaller scales (and higher frequencies) require the consideration of surface tension to adequately resolve, and will consequentially influence all of the coefficients.  In what follows, we develop some simple order-of-magnitude estimates for the case of a circular capillary disk of radius $R$ and mass $m$ that will be evaluated in future work.

By considering a quasi-static interfacial response, one might estimate the effective stiffness $k_e$ arising from buoyancy and capillarity as in equation \ref{eqn:equilibrium2}
\begin{equation}
    k_e=\pi \sigma \left(Bo + 2 \sqrt{Bo}\right)
\end{equation}
where we have further simplified the result assuming ${\mathrm{K}_1(\sqrt{Bo})}/{\mathrm{K}_0(\sqrt{Bo})}\approx 1$ \cite{ho2019direct}. Furthermore, a thin circular disk moving along its axis through a single-phase fluid has an added mass of $\frac{8}{3}\rho R^3$ \cite{lamb1924hydrodynamics}, such that an appropriate estimate for a floating disk might be half of that value 
\begin{equation}
    m_a=\frac{4}{3}\rho R^3.
\end{equation}
(For the equivalent gravity-wave problem, this result is recovered asymptotically in the high-frequency limit \cite{miles1987surface}).  Combining these yields an estimate for the natural frequency of a heaving capillary disk as 
\begin{equation}
    \omega_n^2=\frac{k_e}{m+m_a}=\pi \sigma \frac{Bo + 2 \sqrt{Bo}}{m+\frac{4}{3}\rho R^3}=\frac{3\pi}{4} \frac{Bo + 2 \sqrt{Bo}}{t_{\sigma}^2(m^*+1)}
\end{equation}
where $t_{\sigma}=\sqrt{\frac{\rho R^3}{\sigma}}$ is the inertio-capillary timescale and $m^*=m/m_a$ is a non-dimensional mass ratio comparing the disk mass to the added mass.  For a circular disk of similar scale to a capillary surfer \cite{ho2023capillary} ($m=0.026$ g, $R=0.19$ cm) on an air-water interface, this estimate yields a natural frequency of $f_n=\omega_n/2\pi=18$ Hz. For a circular disk of similar scale to the SurferBot \cite{rhee2022surferbot} ($m=2.6$ g, $R=2.2$ cm) we estimate a natural frequency of $f_n=5.3$ Hz.  For context, the typical driving frequency in each of these experiments is significantly higher than these estimated values.  A similar rough estimate could be crafted for pitching motion at the interface, with the combined heaving and pitching motion ultimately responsible for wave propulsion of a rigid body \cite{benham2024wave}.

To obtain an estimate for the wave resistance $C_e$ for a heaving circular capillary disk, we can equate the power dissipated by a harmonic oscillator in steady state to the power radiated to the capillary-gravity waves generated along the perimeter of the disk of length $2\pi R$. Equating these two quantities we find that,
\begin{equation}
    \frac{1}{2}C_e\omega^2 |z|^2 = 2\pi R E c_g,
\end{equation}
where 
\begin{equation}
    E=\left( \frac{1}{2}\rho g + \frac{1}{2}\sigma k^2\right)A^2
\end{equation}
is the energy density of deep-water capillary-gravity waves \cite{longuet1964radiation} and
\begin{equation}
    c_g = \frac{d \omega}{d k} = \frac{ \rho g + 3\sigma k^2}{2\rho\omega}
\end{equation}
is the corresponding wave group velocity. However, to arrive at an expression for $C_e$, we must assume some relationship between the body amplitude $|z|$ and the wave amplitude $A$.  We estimate this here via a simple volume conservation argument. As the disk completes half a vertical oscillation, a fluid volume of approximately $V_d = 2\pi R^2 |z|$ is displaced. The volume of displaced fluid in a wave of amplitude $A$ over a half cycle is approximately $V_w = 4\pi R A/k$. Equating these two volumes yields the desired relation $\frac{A}{|z|} = \frac{kR}{2}$.  Bringing all of the various pieces together, we arrive at an estimate for $C_e$ as
\begin{equation}
    C_e = \frac{4\pi R}{\omega^2}\left( \frac{1}{2}\rho g + \frac{1}{2}\sigma k^2\right)\left(\frac{kR}{2}\right)^2\left(\frac{ \rho g + 3\sigma k^2}{2\rho\omega}\right).
\end{equation}
In the capillary wave limit, this expression reduces considerably to \begin{equation}
    C_e = \frac{3\pi}{4}\rho\omega R^3.
\end{equation}
Evaluated at the natural frequency, this corresponds to a non-dimensional damping ratio of 
\begin{equation}
    \zeta=\frac{C_e}{2\omega_n m_e}=\frac{9\pi}{32}\frac{1}{m^*+1}.\label{eqn:zeta}
\end{equation}
For our estimated capillary surfer parameters ($m=0.026$ g, $R=0.19$ cm), this corresponds to a damping ratio of $\zeta=0.23$.  Equation \ref{eqn:zeta} suggests that $\zeta<9\pi/32\approx0.88$ in general (by considering the limiting case of $m^*\rightarrow 0$), suggesting that the heaving motion of a surfer should behave as an underdamped simple harmonic oscillator (in the absence of viscous effects).

While these are only crude calculations intended to provide first order-of-magnitude estimates, more complete work that couples the potential (or quasi-potential) flow equations to the rigid-body dynamics of a floating capillary-scale object would expand our fundamental knowledge of wave-structure interactions to the capillary regime, while also helping to elucidate the underlying physics associated with wave propulsion at these scales. 

\subsection{Efficiency and Optimization}\label{sec:efficiency}
A natural engineering question that arises when considering mechanisms of propulsion is one of efficiency.  Longuet-Higgins optimistically noted of wave propulsion that ``the ratio of thrust to power expended on the waves is quite advantageous'' \cite{longuet1977mean}.  Borrowing inspiration from aerospace propulsion \cite{SFORZA20171}, one could define a propulsion efficiency as 
\begin{equation}
    \eta_p = \frac{S_{xx} U}{S_{xx} U + Ec_g} = \frac{\chi Ma}{1+\chi Ma}\label{eqn:efficiency}
\end{equation}
which is the ratio of useful work (thrust times speed) to that plus the radiated wave energy, here assuming waves are only radiated in the wake of the object (as an upper bound).  A similar expression was proposed by Benham et al. \cite{benham2024wave}, however some of the assumptions differ.  The factor $\chi=S_{xx}/E$ is the ratio of radiation stress to energy density (3/2 for the case of pure capillary waves) and $Ma=U/c_g$ is a Mach number that describes the ratio of the surfer speed to the group speed of the waves.  We expect the estimate in equation \ref{eqn:efficiency} to hold for small Mach numbers.  For the case of the SurferBot \cite{rhee2022surferbot}, $Ma\approx 0.03$, leading to an upper-bound efficiency estimate of $\eta_p\approx0.04$.  If one updates the denominator of equation \ref{eqn:efficiency} to assume wave energy is also radiated from the front of the SurferBot with the same amplitude, the estimate reduces to 0.02, in agreement with the corresponding estimates from 2D simulations \cite{benham2024wave}.  As is evident from the image in figure \ref{fig:propulsion}(c), waves are also emitted laterally, and thus the actual efficiency in 3D is likely even lower.  Nevertheless, equation \ref{eqn:efficiency} suggests improved efficiency for higher speeds, corresponding to increasing thrust and reducing drag.  Thrust could be increased by optimizing the kinematics or strengthening the driving.  Longuet-Higgins notes limitations on increased driving in practice however, including the actual maximum steepness of waves as well as avoiding generation of cross waves \cite{longuet1977mean}. \edit{The relevant principles are also expected to shift for supercritical ($Ma>1$) wave propulsors, such as for the case of gunwale bobbing \cite{benham2022gunwale}.}

A first attempt at optimizing wave-driven propulsion was presented by Benham et al. \cite{benham2024wave}, demonstrating that for the case of the SurferBot the efficiency could be increased by more than an order of magnitude (up to $\approx 0.4$) by optimizing the driving frequency and motor positioning.  In particular, the simulations suggest that operation at significantly lower frequencies (on the order of 10 Hz) would improve performance, presumably bringing the forcing frequency closer to the rigid-body natural frequencies of the floater (estimated to be of a similar order in section \ref{sec:WSI}).  Development of a more modular SurferBot in future work where both the motor positioning and frequency can be varied will enable experimental testing of the predictions.  Even more recently, optimal control theory was applied to a further reduced 2D model of wave-driven propulsion in shallow water, \edit{and enabled analysis of both the subcritial ($Ma<1$) and supercritical ($Ma>1$) regimes} \cite{o2025achieving}. A notable conclusion of the work, consistent with our physical understanding of wave propulsion at this point, was that minimizing waves emitted ahead of the body improve performance.  A more complete fundamental understanding of the wave-structure interaction problem will help inform future design changes necessary to achieve such a goal.  \edit{Additionally}, development of high-fidelity 3D models and optimization over larger sets of parameters enabled by modern machine learning tools will be essential towards advancing the state-of-the-art.  Furthermore, it is possible that swarms of wave-driven objects may show improved performance over isolated propulsors by favorably interacting with each others wavefield, as is the case for human swimmers \cite{bolon2023drafting} and ducklings \cite{yuan2021wave}.

\subsection{Robotics}

A number of small-scale interfacial autonomous robots (sometimes referred to as ``microrobots'') have been developed recently for applications such as surveillance, exploration, wildlife imaging, environmental monitoring, and pedagogy \cite{hu2010water,yuan2012bio,chen2018controllable,ko2022high,hou2023light,hartmann2025highly}.  A vast majority of such robots use periodic actuation mechanisms (often inspired by water-walking biological counterparts), and thus emit surface waves during locomotion.  Prior to the SurferBot, it had only been suggested that waves might play a role in propulsion for these devices \cite{lee2019milli} but had not yet been clearly established.  The role of wave momentum still remains largely unaddressed in the most recent literature.  The current SurferBot has an achiral polar design and operates at a fixed frequency.  By leveraging our more advanced understanding of wave-driven torques outlined in section \ref{sec:propulsion}, it may be possible to enable remote steering of a chiral SurferBot via frequency modulation, as was demonstrated on the vibrating bath \cite{barotta2023bidirectional}.  If successful, this would enable both propulsion and steering via a single inertial actuator that need not be in contact with the water.  Furthermore, we have demonstrated that the propulsion modalities are greatly expanded when multiple surfers assemble (section \ref{sec:interaction}), and thus an actively reconfigurable assembly of devices could broaden the tasks an individual is able to perform \cite{saintyves2024self}.

While the SurferBot relies on rigid-body dynamics, many lightweight small-scale robots are inherently compliant.  In fact, carefully designed modal properties of such structures can directly enable propulsion and more complex navigation modalities such as turning \cite{suzuki2007water,becker2013amphibious,lee2019milli}. Recent experimental work on centimeter-scale magnetically forced elastic sheets and untethered compliant autonomous robots have helped elucidate the relevance of traveling waves along the body to efficient interfacial propulsion \cite{wang2022miniature,ren2024undulatory,hartmann2025highly}.  Additional experimental work as well as modeling efforts should be conducted in order to further establish the role of body flexibility on wave-driven propulsion specifically.

Lastly, the emitted waves might serve practical uses beyond self-propulsion.  As discussed in the prior section, swarms of robots might benefit energetically by interacting with their neighbors' radiated wavefields.  Reflected waves might be used for echolocation in complex or crowded environments, a mechanism previously conjectured for some walking-walking insects \cite{voise2014echolocation}.  The forces associated with both surface waves and flows have been shown to be a versatile and controllable tool for manipulating or trapping passive objects at the fluid interface \cite{punzmann2014generation,sherif2020hydrodynamic}, and could be leveraged in a robotic device for collecting samples or transporting cargo analogous to similar tasks performed by active colloids \cite{bishop2023active}.

\subsection{Active Matter and Collective Motion}

        \begin{figure}
        \centering
        \includegraphics{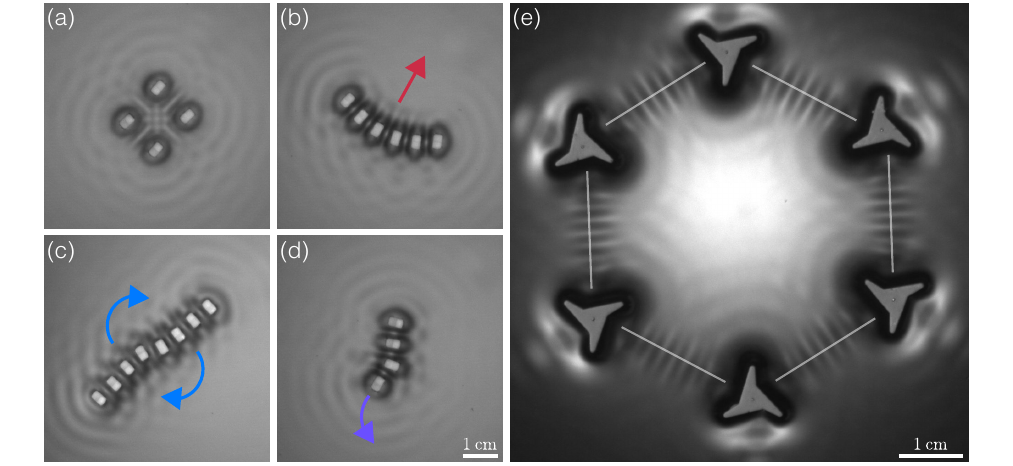}
        \caption{Collections of multiple surfers self-assemble to form collective states that (a) are immobile, (b) move along a straight path, (c) rotate about their center of mass, (d) and move on curved arcs.  (e) Six spinners are magnetically pinned in a regular hexagonal pattern, eventually stabilizing in a frozen configuration where each spinner has cooperatively set up a standing wave with its nearest neighbors.}
        \label{fig:collections}
    \end{figure}

        An active particle is a one that can take in energy, either internally generated or from the surrounding environment, and convert it into directed motion through self-propulsion. Self-propelled particles can be further categorized into three classes: (1) polar particles that translate, (2) spinners that rotate, and (3) chiral particles that are capable of both translation and rotation \cite{liebchen2022chiral}. As reviewed in section \ref{sec:propulsion}, our vibrating bath surfer system is able to realize these three forms of active particles through design of the object's geometry, with additional control possible via modulation of the driving parameters. The resultant collective behavior of self-propelling agents as they interact with one another and their environment constitutes the study of active matter \cite{bechinger2016active, marchetti2013hydrodynamics, elgeti2015physics, cates2015motility, bowick2022symmetry}. Moreover, the type of active particle i.e. polar vs. chiral, influences the collective behavior \cite{liebchen2017collective}.  Active matter research covers many scales, self-propulsion mechanisms, interaction rules, and varying levels of detail in modeling efforts. From phenomenological models of flocking such as the Vicsek model \cite{vicsek1995novel} to learned hydrodynamic equations from simulations and experiments \cite{supekar2023learning}, an effort to better understand the rich dynamics of the collective takes many forms. A few examples of emergent collective behaviors are the formation of polar bands in populations of Quincke rotors \cite{bricard2013emergence}, edge currents in spinning dimers \cite{van2016spatiotemporal}, motility-induced phase separation in swarmalators \cite{adorjani2024motility}, global rotation reversals in air-levitated granular spinners \cite{workamp2018symmetry}, and schooling in populations of fish \cite{zampetaki2024dynamical, filella2018model}, to name a few.

    Synthetic systems have been useful in probing fundamental questions in collective behavior given their tunability. Many of these systems occur at small scales, such as active colloids (e.g. \cite{bricard2013emergence}), wherein inertial effects can be neglected.  However macroscopic experimental systems such as those with granular disks \cite{scholz2018rotating,deseigne2010collective, briand2016crystallization, scholz2018inertial}, brainbots \cite{noirhomme2025}, hexbugs \cite{dauchot2019dynamics} (see Baconnier et al. \cite{baconnier2025self} for a review on self-aligning polar active matter), and our wave-driven surfers are distinctly tunable and allow one to probe the effects of particle inertia on the emergent behaviors. Less is known about the role of fluid inertia as well, and our platform allows for the systematic investigation of collective behavior at intermediate Reynolds numbers \cite{klotsa2019above}.

     In our system, we have demonstrated that two polar or spinner interacting agents can lead to multi-stable dynamic and static states, a feature shared with other mesoscale intermediate Reynolds number systems \cite{chen2025self, dombrowski2022pairwise, gelvan2024hydrodynamic}.  While our work has predominantly focused on pairwise interactions to this point, some preliminary observations of larger collections have been documented with the system (Figure \ref{fig:collections}, \cite{ho2023capillary,oza2023theoretical,sungar2025synchronization}). While systematically moving to larger collections will inevitably yield interesting results, we anticipate that moving to even a three particle system will add considerable richness to the problem, as has been the case for other systems discussed below.  For the case of acoustic interactions, the three-body problem can yield emergent activity \cite{king2025scattered} due to multi-body and non-reciprocal interactions \cite{wu2025non,king2025scattered}. For camphor ribbons, chimera states (the coexistence of coherent and incoherent states in populations of homogeneous oscillators) were observed in a minimal network of three ribbons positioned in an equilateral triangle \cite{sharma2021chimeralike}.  Additionally, in groups of fish, even just three Zebrafish were sufficient to observe multiple facets of collective behavior in the form of schooling, milling, and swarming \cite{zampetaki2024dynamical}. Thus there is evidence from other systems that the three-body problem contains the minimal ingredients to realize new collective behaviors, and \edit{may therefore} represent a natural next step for our system.

    Thinking broadly, underdamped self-propelled particles can be modeled using the framework of active Langevin motion \cite{lowen2020inertial}.  In 2D, this consists of modeling each particle's motion using three degrees of freedom (as in equations \ref{eqn:translation} and \ref{eqn:rotation}): position $\mathbf{r} = (x, y)$ and orientation $\theta$.  A relatively general formulation might thus be posited as
    \begin{align}
        m\ddot{\mathbf{r}} + D\dot{\mathbf{r}} &= D v_0\hat{\mathbf{n}} + \mathbf{F}_{\text{i}} + \mathbf{F}_{\text{e}}+ \mbox{\boldmath$\xi$}_{\text{T}}, \\
        I\ddot{\theta} + D_R\dot{\theta} &= D_R \Omega+\kappa \left(\hat{\mathbf{n}}\times\mathbf{\dot{r}}\right)\cdot \hat{\mathbf{z}} + \mathbf{\tau}_{\text{i}} + \mathbf{\tau}_{\text{e}} +\mathbf{\xi}_{\text{R}}.
    \end{align}
    The self-propulsion speed $v_0$ is modeled as pointing in the direction of the normal vector $\mathbf{\hat{n}} = \left(\cos\theta,\sin\theta\right)$ and a persistent self-propulsion angular velocity is given by $\Omega$. The particles are able to interact with one another $(\mathbf{F}_{\text{i}},\tau_{\text{i}})$ and their environment $(\mathbf{F}_{\text{e}},\tau_{\text{e}})$ in the presence of noise (${\mbox{\boldmath$\xi$}}_{\text{T}},\xi_{\text{R}}$). The term $\kappa \left(\hat{\mathbf{n}}\times\mathbf{\dot{r}}\right)$ corresponds to a self-aligning torque that appears when the center of friction is not aligned with the center of mass of a self-propelling particle \cite{baconnier2025self}, with the sign of $\kappa$ corresponding to whether the particle is an ``aligner'' or a ``fronter'' \cite{ben2023morphological}.  
    Such an approach has been useful in modeling granular spinners in simulation \cite{nguyen2014emergent}, later realized in a tabletop setup \cite{scholz2018rotating}.  In the limit that the inertial timescales are much shorter than any other characteristic timescales in the problem (i.e. $m/D \ll t_c$ and $I/D_R \ll t_c$), we arrive back at overdamped dynamics.  Overdamped dynamics are successfully used to model small-scale systems such as microswimmers and active colloids, but also sometimes applied to larger systems such as collections of whirligig beetles \cite{devereux2021whirligig}. 

    \begin{figure}
    \centering    
    \includegraphics{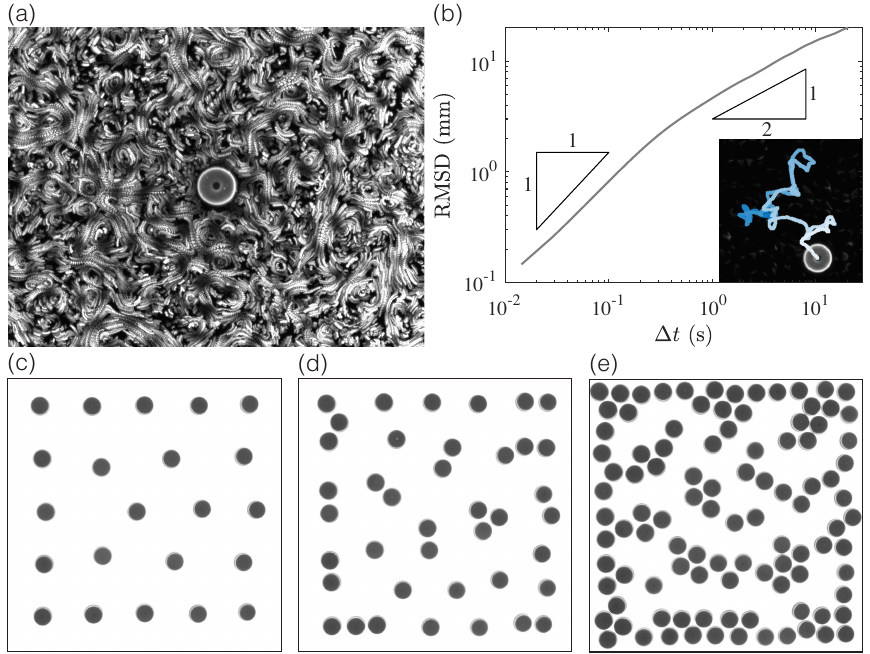}
    \caption{(a) A capillary disk of radius \edit{$R=0.35$} cm moves erratically on the surface of a fluid bath vibrating at $f=140$ Hz, forced by unsteady Faraday waves and the concomitant surface flows. (b) The root-mean-square-displacement of the disk as a function of lag time shows a transition from ballistic to diffusive motion, as in prior work \cite{welch2014ballistic,xia2019tunable}, with a crossover time occurring at $\Delta t\approx 0.5$ s. Inset: sample trajectory of the erratically driven capillary disk. (c-d) Capillary disks with embedded magnets exhibit pattern formation due to competing attractive (capillary) and repulsive (magnetic) interactions \cite{hooshanginejad2024interactions}. (c) Repulsive lattice, (d) cluster, and (e) stripe patterns are observed at different packing fractions in square confinement.}
    \label{fig:NoisePatterns}
\end{figure}

One key element of the active Langevin model that we have thus far neglected is noise. In free space and in the absence of experimental perturbations, our particles move along prescribed paths with no real mechanism to change their trajectory. In other words, this could be described as having an infinite persistence length.  To make better analogy with other systems, it would be useful to understand the influence of noise in our system. One way this might be accomplished in the lab is through the use of supercritical Faraday waves.  Faraday waves spontaneously arise on a vibrated fluid bath above a critical forcing amplitude \cite{faraday1831xvii}. For driving amplitudes just above the threshold, the Faraday waves tend to be ordered as standing waves in specific patterns \cite{edwards1994patterns}. However, further above the threshold, Faraday waves move chaotically while driving unsteady flows, a phenomenon that has been previously used to mimic a thermal bath (\cite{welch2014ballistic}, Figure \ref{fig:NoisePatterns}(a)).  A particle on a Faraday wave-laden bath moves ballistically at short times and diffusively at long times, characterized by its root mean squared displacement (Figure \ref{fig:NoisePatterns}(b)). The effective diffusion constant depends sensitivity on the object size as compared to the Faraday wavelength \cite{xia2019tunable}. By conducting our experiments in the presence of Faraday waves, the Faraday waves may act as white noise with a calculable effective temperature and diffusion constant \cite{welch2014ballistic}. It has been recently shown that macroscopic spheres interacting on a Faraday wave-laden bath are able to undergo structural rearrangements due to the forcing from the unsteady waves \cite{thomson2023nonequilibrium}, acting similarly to colloids in a thermal bath \cite{perry2015two} and assemblies of levitated granular matter \cite{lim2019cluster}.  The inclusion of Faraday waves may introduce other effects beyond noise however, as for instance an asymmetric object can harness energy from the Faraday waves and their associated surface flows, another means for achieving surface propulsion \cite{yang2019passive, francois2018rectification}

    Other interaction forces can also be introduced beyond capillary attraction and wave interaction.  With a slight modification to our 3D-printed molds, we can add small permanent magnets to our floaters \cite{hooshanginejad2024interactions}. Alternatively, magnetic dipoles can be induced in ferromagnetic floating particles by placing them in a uniform magnetic field \cite{vandewalle2012symmetry}.  A vertically-polarized magnetic dipole in each floater introduces a purely repulsive force which scales as $F \sim 1/r^4$. In the absence of waves, magnetic repulsion can balance capillary attraction leading to finite equilibrium spacings in certain regimes \cite{vandewalle2012symmetry,hooshanginejad2024interactions}.      
    Given the competing attractive and repulsive interactions, the equilibrium system can give rise to large-scale pattern formation wherein the particles spontaneously form bubbles, clusters, stripes, and labyrinths depending on the packing fraction (\cite{hooshanginejad2024interactions}, Figure \ref{fig:NoisePatterns}(c-e)).  This pattern formation behavior is characteristic of short-range attractive long-range repulsive (SALR) systems, previously documented at the micro- and nano-scales \cite{royall2018hunting}.  The introduction of waves would introduce an additional competing interaction to the system, with an intrinsic length scale set by the wavelength as discussed in section \ref{sec:interaction}. This new constraint may be at odds with the typical patterns found in SALR systems, leading to entirely new patterns.  Other ways to introduce additional interaction forces between particles might include the use of electric charge \cite{kaneelil2024electrically}.

    Finally, while the studies thus far have focused on particles interacting with each other in free space, the particles can also interact with their environment. Controllable effects might include introducing finite domains where the particles are confined by bordering menisci \cite{thomson2023nonequilibrium} or magnetic traps \cite{perrard2014self}, or the application of time-dependent magnetic fields \cite{grosjean2019capillary,hooshanginejad2024interactions}. Regular or random substrate patterning could also be achieved via bottom topography or magnet arrays, drawing analogy to phenomena in colloidal systems \cite{reichhardt2016depinning, canavello2025polarization}, microswimmers \cite{volpe2011microswimmers}, or other macroscopic wave-driven particle systems \cite{abraham2024anderson}.    

\subsubsection{Relationship to Other Tabletop Active Systems}

    Given the vast scope of research within the field of active matter, we will not attempt to cover all experimental platforms and analytical approaches in our discussion. Instead, we will direct our focus on three related tabletop systems that share strong similarities with our own, and where we believe mutual insights can be gained in addressing questions of collective behavior. The systems we will discuss are vibrated granular media, Marangoni disks, and acoustically levitated particles.

    The granular system typically involves a tabletop setup with centimeter-scale particles, and can be considered a ``dry'' counterpart to the surfer system described here. Particles can be externally driven, via oscillation of the domain itself as is the case for both polar \cite{deseigne2010collective} and spinner granular active particles \cite{scholz2018inertial, scholz2018rotating}, or internally driven as is the case for hexbugs \cite{dauchot2019dynamics}. For all such particles, the vibration is converted into directed motion via angled legs of the object which exploit translational and/or rotational asymmetry. Alternatively, one can use an ``air-hockey'' table setup to power granular spinners, converting air from the table below into directed motion via prescribed air ducts in the design \cite{workamp2018symmetry, farhadi2018dynamics}. Granular particles are easily manipulated since the particles can be 3D printed, allowing for the geometry to be readily altered.  In fact it was recently suggested that ``...granulate particles will play a leading role as paradigmatic realizations for active matter models. Since they are macroscopic, the particles can more easily be manipulated and tailored'' \cite{lowen2020inertial}.  Our system shares this benefit.  

    There are many similarities between our system and the granular walkers. Both use mechanical vibration as a means for self-propulsion, and the systems both have an accessible tabletop setup that can easily be modified. The bath/arena is another factor that is simple to customize with 3D-printed or laser-cut pieces allowing for systematic control over confinement or substrate effects. While the systems share these strengths, the systems critically differ in the admissible types of interaction. For the granular system, active particles interact with one another via local steric interactions. The collision interactions tend to be modeled using elastic contact forces and are inevitably short ranged. In contrast, our particles interact via relatively long-range interfacial forces which have a range of interaction spanning multiple body lengths. In the absence of oscillation, our particles would naturally attract one another and cluster on the interface via capillary attraction (section \ref{sec:cheerios}). In the presence of waves however, this effect can be stabilized, and interactions tend to occur at quantized distances set by the wavelength \cite{oza2023theoretical, barotta2025synchronization, sungar2025synchronization} as reviewed in section \ref{sec:interaction}. The granular active particle literature is filled with many future directions that may prove fruitful to attempt with our setup. For example, with a single polar active particle, one could design a harmonic trap to draw parallels to the analogous investigation done with a hexbug on a parabolic dish \cite{dauchot2019dynamics}. One may expect that for a wave-mediated system, interactions with the container itself may lead to new emergent behaviors due to wave interference \cite{tarr2024probing}. At the collective level, the consideration of populations of counter-rotating spinners \cite{scholz2018rotating, nguyen2014emergent} will likely yield new results when the constituents interact with all others via their wavefields rather than only fleetingly through pairwise collision.  

    Marangoni surfers also represent a tabletop approach with particles of centimetric scale, that might be viewed as a physicochemical counterpart to the system described herein. The mechanism underlying self-propulsion and interaction is Marangoni stresses which are driven by gradients in the surface tension (in analogy to gradients in the wavefield height in our system).  Marangoni surfers typically feature a fuel source in the form of a chemical such as camphor that is controllably released from the body onto the liquid interface.  Modifications to the geometry of a Marangoni surfer can be used, similar to our wave-mediated setup, to control the trajectory of the objects \cite{sur2021effect, wilt2024activecheerios}. Steering can be realized via active deformation on the floating object coupled with the controlled emission of chemical fuel \cite{huang2024self}. Recent efforts to fabricate Marangoni ``Cheerios'' leverage a setup similar to the granular walkers by using a 3D-printed body for the object, coupled with chemical fuel (water-ethanol mixtures), resulting in a low-cost interfacial propulsor capable of curvilinear trajectories \cite{wilt2024activecheerios}. 
    
    Marangoni surfers have been used in a variety of modular setups to explore questions such as first passage time of an active particle \cite{biswas2020first}, stochastic migrations in a dumbbell configuration \cite{upadhyaya2024stochastic}, trapping via external perturbations of the interface \cite{tiwari2023capture}, and 1D collective behavior analogous to traffic flow in annuli \cite{suematsu2010collective,heisler2012swarming}. The use of camphor ribbons, as opposed to disks, has enabled controlled study of Marangoni spinners \cite{sharma2019rotational}. Here, the camphor ribbons were each pinned on one end and interacted with one another via Marangoni stresses on the interface. This system revealed synchronization phenomena reminiscent of our magnetically pinned spinner setup and has shown the possibility of more exotic states when considering collections of $N=3$ spinners \cite{sharma2021chimeralike}.  While our system shares many similarities as a low-cost interfacial setup with fluid-mediated interactions, our system does not need to incorporate physicochemical considerations into modeling propulsion and interaction.  Furthermore, for the case of Marangoni surfers the background is time dependent, as chemicals are continuously released with the interface eventually becoming fully saturated and arresting all motion.

   Finally, the acoustically levitated system  also features a tabletop approach that may be viewed as a sound-wave counterpart to our system. When an object is placed in an acoustic field, it experiences a net force toward a node or antinode. When multiple objects are placed in an acoustic field, the particles induce additional forces between one another that arise due to wave scattering. For conditions where the particle and particle-particle separations are much smaller than the wavelength of the acoustic wave (i.e. near-field Rayleigh scattering), particles experience an attractive force due to the scattering of sound waves that decays algebraically with distance ($F \sim 1/r^4$) \cite{silva2014acoustic}. At the same time, a streaming flow develops due to the non-zero viscosity of the surrounding medium leading to a repulsive force which decays exponentially in space ($F\sim e^{-r}$) \cite{wu2023hydrodynamic}. As multiple bodies interact in the acoustic field, a host of novel behaviors arise such as emergent propulsion \cite{st2023dynamics} \edit{and} activity in populations of three small spheres \cite{wu2025non, king2025scattered}. \edit{When acoustic interactions are \edit{combined} with other fields, groups exhibit collective phenomena such as melting (due to coupled hydrodynamics) \cite{wu2023hydrodynamic}, pattern formation (due to contact cohesion) \cite{wu2025pattern}, and dynamic self-assembly (due to coupled electrostatics) \cite{shi2025electrostatics}.} (see Lim et al. \cite{lim2024acoustic} for a recent review on acoustic multi-body structure and dynamics). This system inspires new ways of thinking about multi-body interactions across scales. In particular, we hope to use the acoustic-levitated system as inspiration when considering many-body interactions that couple the scattered wavefield to the secondary streaming flows. As shown in King et al. \cite{king2025scattered}, nonreciprocal interactions can arise in problems of wave-scattering where we may expect our interaction force law (equation \ref{eqn:force}) would need to be updated to account for higher-order effects. Given the wealth of interactions we have encountered in our interfacial system where the particles are of comparable size to the wavelength, we anticipate novel behavior for the acoustic setup when not in the near-field Rayleigh scattering limit, i.e. when the particle and particle-particle length scales are comparable to the wavelength.

    Overall, we believe that our system of wave-propelled interfacial particles holds promise as a tabletop experiment to probe questions in collective behavior synergistically with the other setups described.  Furthermore, theoretical models can be developed directly from first principles for our system, reducing the need for fitting parameters, such has already been successfully demonstrated for the pairwise wave-mediated interaction force between oscillating bodies (equation \ref{eqn:force}, \cite{oza2023theoretical, de2018capillary}).

\subsection{Hydrodynamic Quantum Analogs}
The fluid interface has long represented a fruitful source for analogy with other concepts in physics, ranging from the capillary assembly crystal physics of Bragg \cite{bragg1942model,bragg1947dynamical} to the ripple tank experiments used to explain the wave nature of light by Young \cite{young1832bakerian} and the Aharonov-Bohm effect by Berry \cite{berry1980wavefront}.  \edit{Two decades ago}, Couder and Fort discovered that a droplet on a vibrating bath can self-propel along the fluid interface by virtue of its own self-generated wavefield, suggesting analogy with quantum particles as a manifestation of a macroscopic wave-particle duality \cite{couder2005walking,Protiere2006,couder2006single}.  The topic has since grown into a field of its own, with a number of reviews covering the progress \cite{bush2015pilot,bush2020hydrodynamic,bush2024perspectives}, and the list of analogies with quantum phenomena continuing to grow.  The analogies have also been expanded to connect with other topics, such as microscopic spin systems \cite{saenz2021emergent} and nuclear physics \cite{valani2025hydrodynamic}.  The surfer system represents another hydrodynamic pilot-wave system, although it has yet to be explored within the context of quantum analogs.  The walking droplet and surfer systems share some key similarities and differences, with a few highlights outlined below.

The systems rely on a nearly identical experimental platform centered around a vibrating fluid bath \cite{harris2015generating}.  Being driven periodically, the wavefields in both systems are monochromatic, leading to the emergence of spatially quantized interactions with each other and their surroundings \cite{couder2005walking,protiere2006particle,oza2017orbiting,ho2023capillary,oza2013trajectory}.  Furthermore, in companion modeling efforts, both the droplets and surfers are modeled as responding to the interface via interaction with local height gradients \cite{protiere2006particle,molavcek2013drops,oza2013trajectory,oza2023theoretical}.  Nevertheless, there are some notable distinctions between the systems.  While the walking droplets self-propel via a spontaneous symmetry breaking for very restricted combinations of fluid and driving parameters \cite{protiere2006particle,molavcek2013drops}, the surfers symmetry is broken by design, and thus propel for essentially all parameters where waves are generated \cite{ho2023capillary,barotta2023bidirectional}.  The droplets execute a remarkably rich spectrum of nonlinear vertical bouncing dynamics \cite{protiere2006particle,molavcek2013drops,wind2013exotic}, whereas the surfers do not lose contact with the interface and have been documented to only respond at the driving frequency, in an apparently linear manner \cite{ho2023capillary}.  For the case of walking droplets, most experiments are performed just below the Faraday threshold, with an individual droplet's individual wavefield principally described by a standing wave \cite{eddi2011information}.  By operating near the threshold, the longevity of the damped Faraday waves (i.e. ``memory'') can be directly tuned via the driving, a feature responsible for many of the quantum-like behaviors documented with the system.  The surfer experiments are also performed below the Faraday threshold, but the wavefield is an outwardly propagating wave whose influence vanishes on a fixed timescale associated with viscosity \cite{ho2023capillary,oza2023theoretical}.  The surfer platform is significantly less mature however, as it was only recently discovered, and thus the list of differences will surely both grow and diminish with time.  At the same time, other classical wave-particle systems have also recently been realized or proposed, including neutrally buoyant objects exciting internal waves \cite{le2022swimming,gsellmacroscopic} and suspended particles surfing on acoustic fields \cite{roux2022self,martischang2023acoustic}.  \edit{The Faraday wave system itself has also recently been shown to exhibit behaviors analogous to driven Bose-Einstein condensates \cite{liu2024polygonal}.} The future exploration of quantum analogs with these classical particle-wave systems will provide an additional richness and breadth to the field of hydrodynamic quantum analogs, while also enabling the development of new analogies and ultimately providing deeper insights into the interpretations of quantum mechanics.

\section{Conclusion}\label{sec:conclusion}

In this work, we have provided an overview of the recent progress on wave-driven propulsion and interaction of floating bodies, with particular attention to objects at the capillary scale.  Oscillating bodies may self-propel at the interface by virtue of the unbalanced momentum flux associated with their self-generated propagating wavefield.  These waves extend to great spatial extent along the interface, resulting in rich wave-mediated multi-body interactions.  We have further outlined a number of potential future directions that have emerged as a consequence of this ongoing work, spanning across fluid mechanics, engineering, and fundamental physics.  The ideas laid out herein are by no means exhaustive, and are limited only by our own domain knowledge, experience, and creativity.  As such, we encourage others to become involved in this emerging field, with the relative nascency and overall accessibility of the system fostering easy entry points with a high likelihood of discovery.

\begin{figure}
    \centering
    \includegraphics{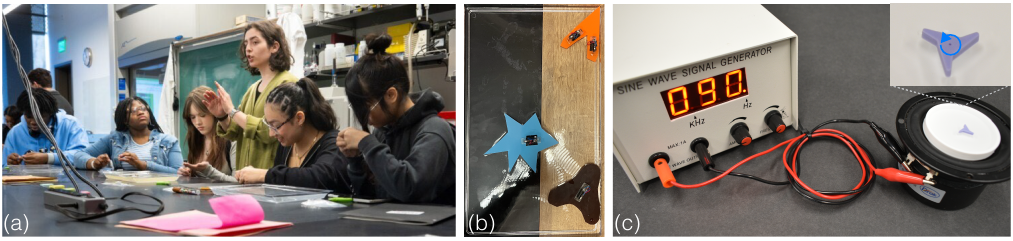}
    \caption{(a) Undergraduate student Yesenia Gomez leads a SurferBot workshop as part of Brown University's 2025 STEM Day outreach program (Photo Credit: Kevin Stacey). (b) Participants are encouraged to customize their SurferBot design during these workshops, exploring the influence of motor placement and geometry on propulsion. (c) A plastic dish adhered to a small sub-woofer is vertically oscillated by a signal generator, driving rotation of a floating capillary spinner.}
    \label{fig:outreach}
\end{figure}

Beyond research interest and technological applications, the key phenomena can be reproduced with low-cost hands-on setups, in a similar spirit to the hexbug in ``dry'' active matter \cite{barona2024playing}.  For instance, we have successful leveraged the SurferBot concept for numerous outreach activities, where K-12 students assemble the device (costing $<\$1$ per unit) while being introduced to concepts such as surface tension, momentum, and waves (Figure \ref{fig:outreach}(a)). Participants are also given the opportunity for ample open-ended exploration. After assembling the base unit, they are encouraged to modify the design and observe the outcome of their decisions, for instance observing changes in speed or propulsion direction with regard to the motor positioning or shape of the base (Figure \ref{fig:outreach}(b)).  This approach also gives the students an opportunity to naturally experience design iteration while encountering engineering tradeoffs, for instance smaller platforms tend to reduce drag (leading to higher speeds), however if too small, the device will sink.  The vibrating bath setup can also be reproduced affordably for use in pedagogy and outreach.  While experimental repeatability certainly benefits from the development and use of a highly controlled vibration setup \cite{harris2015generating}, nearly all of the core phenomena discussed herein can \edit{be} readily observed on a simple setup consisting of a plastic dish adhered to a small sub-woofer (Figure \ref{fig:outreach}(c), \cite{harris2017visualization,ho2022reconfigurable}).  In fact, the bidirectionality of capillary spinners was originally discovered on a similar sub-woofer rig, with refined data collection then executed on a more controlled laboratory setup \cite{barotta2023bidirectional}.  Tractable research questions can also be readily devised and integrated into curricular research experiences, such as in a ``PhyExp'' style course as pioneered by Yves Couder \cite{eddi2020experimental} or in other Course-based Undergraduate Research Experience (CURE) courses or modules \cite{cure}, enabling broad participation in the research experience.

\vspace{0.1in}

The authors recognize the invaluable contributions to this line of work by students, postdocs, and collaborators of the Harris Lab: Ian Ho, Giuseppe Pucci, Eugene Rhee, Stuart Thomson, Robert Hunt, Luke Alventosa, Maya Lewis, Eli Silver, Navid Hooshanginejad, and Anand Oza; as well as financial support from the National Science Foundation (NSF CBET-2338320), the Office of Naval Research (ONR N00014-21-1-2816; ONR N00014-21-1-2670), and the National Defense Science and Engineering Graduate (NDSEG) Fellowship Program.

\bibliography{refs}

\end{document}